\documentclass[fleqn,usenatbib]{mnras}

\usepackage{newtxtext,newtxmath}

\usepackage[T1]{fontenc}

\DeclareRobustCommand{\VAN}[3]{#2}
\let\VANthebibliography\thebibliography
\def\thebibliography{\DeclareRobustCommand{\VAN}[3]{##3}\VANthebibliography}

\newcommand\gaia{\textit{Gaia }}

\usepackage{graphicx}	
\usepackage{amsmath}	
\usepackage{amssymb}	

\usepackage{datatool, filecontents}
\hypersetup{citecolor=blue}
\DTLsetseparator{,}
\DTLloaddb[noheader, keys={thekey,thevalue}]{main_data}{main_data.dat}
\newcommand{\var}[1]{\DTLfetch{main_data}{thekey}{#1}{thevalue}}

\newcommand\solarmass{\ensuremath{\mathrm{M}_{\odot}}}

\newcommand{\red}[1]{{#1}}

\defcitealias{NSS}{Gaia Collaboration, Arenou et al. 2022}

\title[Triage of Gaia astrometric binaries]
{Triage of the {\it Gaia} DR3 astrometric orbits. I. A sample of binaries with probable compact companions}

\author[S. Shahaf et al.]{
S. Shahaf,$^{1}$\thanks{E-mail: sahar.shahaf@weizmann.ac.il}
D. Bashi,$^{2}$
T. Mazeh,$^{2}$
S. Faigler,$^{2}$
F. Arenou,$^{3}$
K. El-Badry,$^{4,5,6}$
and H.~W. Rix$^{6}$
\\
$^{1}$Department of Particle Physics and Astrophysics, Weizmann Institute of Science, Rehovot 7610001, Israel\\
$^{2}$School of Physics and Astronomy, Tel Aviv University, Tel Aviv, 6997801, Israel\\
$^{3}$GEPI, Observatoire de Paris, Université PSL, CNRS, 5 Place Jules Janssen, 92190 Meudon, France\\
$^{4}$Center for Astrophysics | Harvard \& Smithsonian, 60 Garden Street, Cambridge, MA 02138, USA\\
$^{5}$Harvard Society of Fellows, 78 Mount Auburn Street, Cambridge, MA 02138\\
$^{6}$Max-Planck Institute for Astronomy, Königstuhl 17, D-69117 Heidelberg, Germany}

\date{Accepted XXX. Received YYY; in original form ZZZ}

\pubyear{2022}

\begin{document}
\begin{filecontents*}{main_data.dat}
N_initial,127026
N_Halbwachs_filter,126624
N_TI_filter,110401
N_with_ageFeh,49059
BH_alpha,10
N_class3,177
N_bh,8
N_ns,24
c3_nss_and_now,148
c3_nss_not_now,433
c3_not_nss_and_now,29
c3_nss_clean,581
n_wd_stripe,93
n_ns_stripe,24
n_bh_stripe,8
n_interm_stripe,48
n_elm_stripe,48
N_period_filter,101380
\end{filecontents*}

\label{firstpage}
\pagerange{\pageref{firstpage}--\pageref{lastpage}}
\maketitle

\begin{abstract}
In preparation for the release of the astrometric orbits of {\it Gaia},  Shahaf et al.~(2019) proposed a triage technique to identify astrometric binaries with compact companions based on their astrometric semi-major axis, parallax, and primary mass. The technique requires the knowledge of the appropriate mass-luminosity relation to rule out single or close-binary main-sequence companions. The recent publication of the \textit{Gaia} DR3 astrometric orbits used a schematic version of this approach, identifying $735$ astrometric binaries that might have compact companions. In this communication, we return to the triage of the DR3 astrometric binaries with more careful analysis, estimating the probability for its astrometric secondary to be a compact object or a main-sequence close binary. We compile a sample of $\var{N_class3}$ systems with highly-probable non-luminous massive companions, which is smaller but cleaner than the sample reported in \textit{Gaia} DR3. The new sample includes $8$ candidates to be black-hole systems with compact-object masses larger than $2.4$ \solarmass. The orbital-eccentricity--secondary-mass diagram of the other $169$ systems suggests a tentative separation between the white-dwarf and the neutron-star binaries. Most white-dwarf binaries are characterized by small eccentricities of about $0.1$ and masses of $0.6$ \solarmass, while the neutron star binaries display typical eccentricities of $0.4$ and masses of $1.3$ \solarmass. 
\end{abstract}

\begin{keywords}
astrometry -- binaries: general -- stars: white dwarfs -- stars: neutron -- stars: black holes
\end{keywords}



\begin{figure*}
	\includegraphics[width= 2.0\columnwidth,trim={2cm 0 2cm 0},clip]{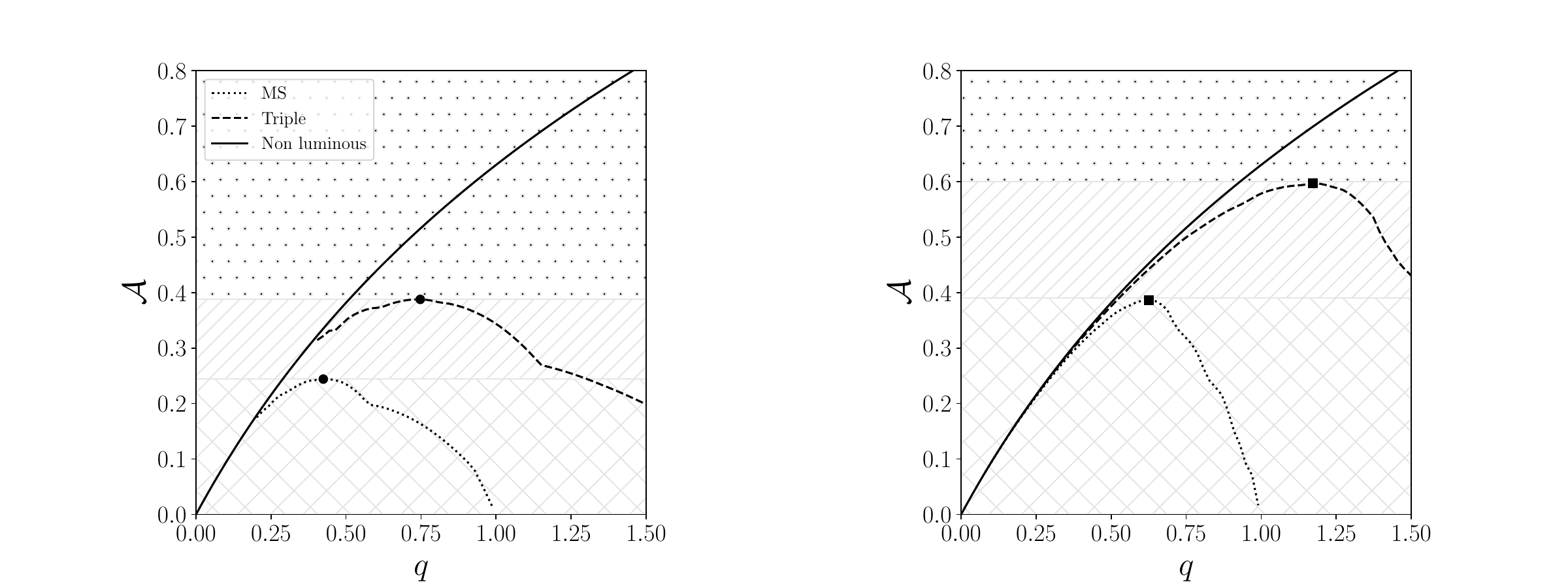}
    \caption{Expected AMRF curves as a function of $q$ for primaries of $0.4 \, \solarmass$ (left) and $1.0\, \solarmass$ (right), based on the empirical MLR of \citet{mamajek13} for the {\it Gaia} G band. Lines represent three limiting cases for an astrometric companion: a single MS secondary is shown as a dotted line; a close equal-mass MS binary, as the astrometric companion, appears as a dashed line, and a non-luminous companion is plotted as a solid line. No binary can exist below the dotted line. All triple systems have to reside between dashed and dotted lines. Binaries above the dashed lines must be compact objects. Points mark the maximum  $\mathcal{A}$ for single and close-pair MS companions. We do not know the actual mass ratio, so the maximum points divide the possible $\mathcal{A}$ values into three ranges.} 
    \label{fig:amrf_curve1}
\end{figure*}

\section{Introduction}
\label{introduction}

The population of binaries with white-dwarf (WD), neutron-star (NS), or black-hole (BH) companions is of great interest. It sheds light on the properties of the binaries for which the more massive primary component completed its main-sequence (MS) phase and on the dramatic processes accompanying the transition into a compact object \citep[e.g.,][]{heger03, cerda18}. Astrometry is an important tool to study this population, as it is sensitive to binaries with orbital periods of the order of a few years, depending on the binary distance, which corresponds to orbital separations to which other techniques, spectroscopy or photometry, are less sensitive \citep[e.g.,][]{jorissen08}. Furthermore, unlike spectroscopic binaries for which the orbital inclination is not known, 
the compact-object mass in astrometric binaries can be determined, and the three types of compact objects can, in principle, be distinguished \citep[e.g.,][]{halbwachs22}. 

The \gaia astrometric space mission \citep{gaia16} provides a promising detection channel, as it is expected to detect an unprecedentedly large number of astrometric binaries. For example, theoretical studies predict that the \textit{Gaia} mission carries the potential of discovering hundreds of binaries with non-interacting BHs in orbital periods  ${\lesssim}\,5$ years \citep[e.g., ][]{breivik17, mashian17, yamaguchi21, janssens22, chawla22}. NSs and WDs should be even more frequent \citep[e.g.,][]{fryer12}. Note, however, that the stringent selection criteria imposed on the DR3 sample of astrometric binaries \citep[see ][]{halbwachs22}, designed to reduce the contamination of the astrometric catalogue by spurious signals, probably excluded many of these systems, impairing the detection of the compact objects.

In preparation for the release of the astrometric orbits of {\it Gaia}, \citet*[][]{shahaf19} proposed a triage technique to identify astrometric binaries that have compact companions based on their derived semi-major axis, parallax, orbital period and the estimated primary mass. The technique requires the knowledge of the proper mass-luminosity relation (MLR) to rule out a single or a close-binary MS companion. Indeed, the DR3 binary release \citepalias[][hereafter {NSS} --- Non-Single Stars]{NSS} used a schematic version of this approach, with the MLR of \citet{mamajek13}, to identify $735$ astrometric binaries with compact companions \citep[see also][for a different approach]{andrews19, andrew22}.
\defcitealias{NSS}{NSS}

In this communication, we return to the triage of the DR3 astrometric binaries with a more careful analysis that uses a more conservative  MLR to identify compact-secondary binaries based on a suit of MIST isochrone grids.\footnote{MESA Isochrones \& Stellar Tracks. See \href{https://waps.cfa.harvard.edu/MIST/}{waps.cfa.harvard.edu/MIST/.}} 
We derive a less contaminated catalogue of compact companions, identifying astrometric binaries with WD, NS or BH companions.

The paper is structured as follows: In Section~\ref{sec:triage} we briefly describe our astrometric triage scheme and discuss the effect of stellar age and composition on this technique. In Section~\ref{sec: triage sample} we present the triage of \textit{Gaia} DR3 binaries, provide a list of class-membership probabilities, and compile a sample of systems that are very likely to host a compact object in Section~\ref{sec: c3}. In Section~\ref{sec: m2-e} we show some of the emerging properties of our compact-object sample. Finally, in Section~\ref{sec: discussion} we briefly discuss the sample and preliminary findings and propose some ideas for future work.

\section{Astrometric Triage}
\label{sec:triage}

\begin{figure*}
    \centering
    \begin{minipage}{0.475\textwidth}
        \centering
        \includegraphics[width=0.9\textwidth]{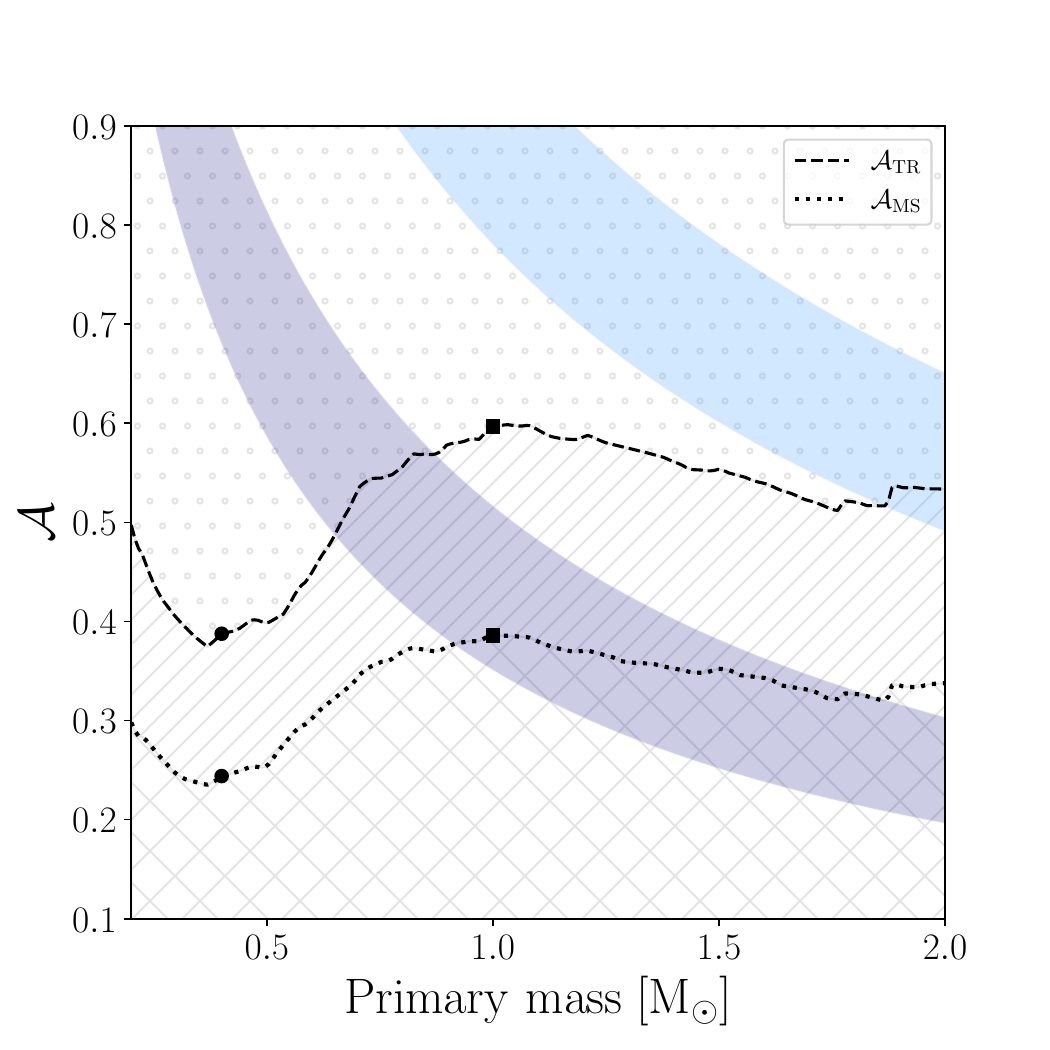} 
        \caption{Maximum $\mathcal{A}$ for an MS secondary (dotted line) and an MS close-binary companion (dashed line) as a function of primary mass, based on \citet{mamajek13} observed MLR. Black circles and squares show the maximum values of Fig.~\ref{fig:amrf_curve1}. Purple and light-blue stripes illustrate the expected $\mathcal{A}$ of binaries with WD, at $0.45{-}0.75\,\solarmass$, and NS, at $1.4{-}2.1\, \solarmass$ companions, respectively. The binary position does not depend on the period or the parallax of the system. The figure suggests that some white-dwarf and most NS binaries are expected to reside above the corresponding $\mathcal{A}_{\textsc{tr}}$, and therefore can be identified as having compact companions.}
    \label{fig:amrf_curve}
    \end{minipage}\hfill
    \begin{minipage}{0.475\textwidth}
        \centering
        \includegraphics[width=0.9\textwidth]{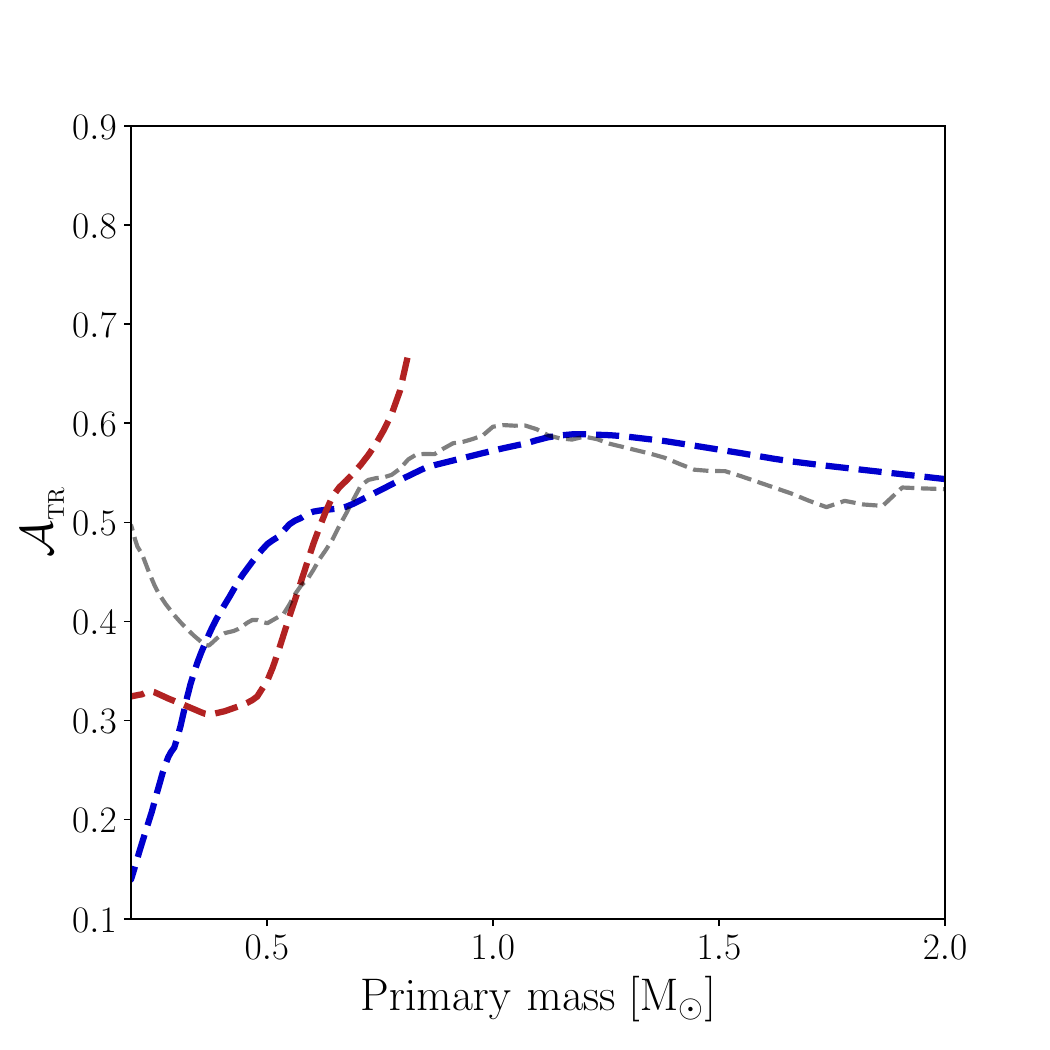} 
        \caption{
        Class-III limiting value, $\mathcal{A}_{\textsc{tr}}$, as a function of the primary mass for two different populations. Dashed red and blue lines represent the MIST-based limits for old and young stellar populations, respectively. Grey dashed line represents the limit obtained from the empirical MLR of \citet{mamajek13} (dashed), which was used by \citetalias{NSS}. 
        Note that the range of the old-population curve ends at $\sim0.9$ $\solarmass$, for which the corresponding primaries are expected to leave the main sequence.\newline}
    \label{fig:MLR_curve}
    \end{minipage}
\end{figure*}

In this section, we re-discuss the astrometric triage introduced by \citet[][]{shahaf19}, with a focus on using the appropriate MLR for MS stars.

Consider an astrometric binary with an angular semi-major axis $\alpha_0$. For an unresolved binary, $\alpha_0$ reflects the motion of the centre-of-light around the binary centre-of-mass. In cases where the more luminous primary star is significantly brighter than its secondary companion, the photo-centre of the system is located near the primary star's position. This could happen, for example, if the secondary is a faint sub-stellar companion or a compact object. On the other hand, if both components are luminous,  the photo-centre is located near the centre-of-mass of the binary, up to a point where no astrometric orbit can be detected.

\citet{shahaf19}  presented the astrometric mass ratio function (AMRF),
\begin{equation}
    \mathcal{A} \equiv \frac{\alpha_0}{\varpi} \bigg(\frac{M_1}{\textrm{M}_\odot}\bigg)^{-1/3} \bigg(\frac{P}{\textrm{yr}}\bigg)^{-2/3}\, , 
    \label{eq: AMRF 1}
\end{equation}
where $P$ and $\varpi$ are the orbital period and parallax, \red{and  $M_1$ is the mass of the primary, more luminous, star.} $\mathcal{A}$ can be determined for every astrometric binary for which  $M_1$ is known.

The unknown mass ratio $q=M_2/M_1$  is linked to AMRF via
\begin{equation}
    \mathcal{A}= \frac{q}{(1+q)^{2/3}}\,\bigg(1 - \frac{\mathcal{S}(1+q)}{q(1+\mathcal{S})} \bigg)\, ,
    \label{eq: AMRF 2}
\end{equation}
\red{
where $\mathcal{S}=F_2/F_1$ is the luminosity ratio between the two components. The term inside the parenthesis in equation~(\ref{eq: AMRF 2}) accounts for the fact that we follow the orbit of the photo-centre, rather than that of the primary star (see, for example, \citealt[][]{kamp75}); assuming the mass-luminosity relation is superlinear, $\mathcal{S}\leqslant q$ and this term is non-negative. 
}

\citet[][]{shahaf19} showed that whenever the luminosity ratio of the two possible MS stars can be expressed as a function of the mass ratio, $\mathcal{S}(q)$, one is able to place some constraints on the nature and properties of the faint companion in the binary system. 
\red{
A lower estimate for the mass ratio is obtained by assuming that the secondary companion is non-luminous, namely, plugging $\mathcal{S}=0$ into equation~(\ref{eq: AMRF 2}). Under this assumption, the AMRF is a function of the mass ratio alone.}

\red{
The minimal mass ratio, $q_{\min}$ is a root of the polynomial $\mathcal{A}^{-3}q^3 -q^2 - 2q -1$. Since $\mathcal{A}$ is a positive number, the minimal mass ratio is unique and can be obtained analytically \citep[e.g.,][]{heacox95, shahaf17, andrew22}. 
The corresponding  minimal secondary mass, i.e., the mass of the companion assuming that it does not emit light, is given by
\begin{equation}
    M_{2,\min} = q_{\min} \cdot M_1.
\end{equation}
This lower limit on the mass may, in many cases, constrain the compact object's nature. However, this can only be done if we rule out the possibility of a single MS secondary or a companion who is by itself a close MS binary. 
}

\subsection{AMRF classification}

To illustrate the triage approach, we plot in Figure~\ref{fig:amrf_curve1} two theoretical AMRF curves as a function of the mass ratio $q$, for $0.4$ and $1.0$~$\solarmass$ MS primary stars. The dotted (lower) curves in the two panels present binaries with a single MS secondary; the dashed (upper) curves triple systems, with a close equal-mass MS binary as the astrometric secondary; and the solid curves binaries with a non-luminous companion. The dotted and dashed lines were derived with the \textit{Gaia} G-band MLR of \citet{mamajek13}.

Binaries with MS companions have to reside on the lower dotted curves, with a position that depends on the mass ratio of the astrometric binary. Triple systems with close-binary MS companions could be located anywhere below the upper dashed line, depending on the mass ratio of the close binary. Only triple systems with equal-mass close-binary companions have to be on the upper dashed line, with a position that depends on the wide-binary mass ratio. Binaries with compact companions have to reside on the continuous curve.
\red{Note that the positions of 
realistic systems do not necessarily fall on the expected position.  This is because the shape of the distinguishing lines is affected by the accuracy of the assumed MLR, and because a binary position on the diagram is affected by the measurements uncertainties.}

The AMRF theoretical curves have maximal values --- $\mathcal{A}_{\textsc{ms}}$ for the single MS secondary and $\mathcal{A}_{\textsc{tr}}$ for the triple-system curve. These values, which depend on the primary mass, are noted in the figure by dots and squares. Any wide binary  with $\mathcal{A}>\mathcal{A}_{\textsc{tr}}$ probably has a compact companion. If the companion is a single object, the system has to reside on the continuous curve. In such a case, the companion mass can be derived from the value of $\mathcal{A}$.

In the general case, though, one cannot determine the value of $q$, even if the primary mass is known, because the luminosity of the secondary is unknown. Therefore, to identify unresolved astrometric binaries that are likely to host a compact object as their faint companion, \citet{shahaf19} divided the astrometric binaries into three classes, based on their measured AMRF value, $\mathcal{A}$:
%
\begin{enumerate}
\item \textit{Class-I binaries} ($\mathcal{A}<\mathcal{A}_{\textsc{ms}}$), where the companion is most likely a single MS star. The class-I parameter space is shown as a crisscrossed area in Figure~\ref{fig:amrf_curve}. \vspace{2.5pt}
\item \textit{Class-II binaries} ($\mathcal{A}_{\textsc{ms}}<\mathcal{A}<\mathcal{A}_{\textsc{tr}}$), where the companion cannot be a single MS star, but can be either a MS close binary or a compact object. The class-II parameter space is denoted with slanted lines. \vspace{2.5pt}
\item \textit{Class-III binaries} ($\mathcal{A}>\mathcal{A}_{\textsc{tr}}$), where the companion cannot be a single MS star nor a close MS binary; these systems are likely to host a compact object secondary. The class-III parameter space is highlighted by small circles.
\end{enumerate}

Figure~\ref{fig:amrf_curve1} demonstrates that the limiting values $\mathcal{A}_{\textsc{ms}}$ and $\mathcal{A}_{\textsc{tr}}$ vary as a function of $M_1$. Figure~\ref{fig:amrf_curve} shows the limiting AMRF values as a function of the primary mass. We added in Figure~\ref{fig:amrf_curve} purple and light-blue stripes that illustrate the expected $\mathcal{A}$ values of WDs, at $0.45{-}0.75\,\solarmass$, and NSs, at $1.4{-}2.1\, \solarmass$.

Note that the locations of binaries with primaries ${\lesssim} 2$ $\solarmass$ and NS companions
are all in the class-III region, 
making their identification relatively simple. The WD stripe, on the other hand, is only partially above the $\mathcal{A}_{\textsc{tr}}$ curve. This implies that only binaries with massive WDs can be identified as such, while  binaries with low-mass WDs and relatively massive primaries will escape detection. We will come back to this point in an accompanying paper.

\subsection{Re-consideration of class-II and class-III limits}
\label{sec: age metallicity based limits}
The AMRF limits depend on the assumed MLR, which in turn depends on the age and chemical composition of the specific binary. The observed MLR of \citet{mamajek13} used in Figure~\ref{fig:amrf_curve1} and \ref{fig:amrf_curve} is an averaged relation, taken over the distribution of ages and compositions in the Solar neighborhood. While this relation can properly describe the population of stars in the field, this is not necessarily the case when considering two stars in a particular binary system. Assuming the two stars were formed at the same time and have the same composition, their relative flux contribution follows some specific isochrone track rather than the local averaged MLR.

To demonstrate this point Figure~\ref{fig:MLR_curve} presents two $\mathcal{A}_{\textsc{tr}}$ curves for two different populations: a young population, with age of $126$ Myr and $[{\rm Fe/H}]=0.5$ (a dashed-blue curve), an old population of $12.6$ Gyr and $[{\rm Fe/H}]=-2.5$ (dashed-red curve). To derive the first two curves we simulated a synthetic stellar population using \texttt{ArtPop} package\footnote{See the online documentation at \href{https://artpop.readthedocs.io/en/latest/}{artpop.readthedocs.io}} \citep{greco21} and the MIST isochrone grids. The figure also displays the (upper) curve of Figure~\ref{fig:amrf_curve1} (dashed-gray curve), based on  \citet{mamajek13} MLR, which was used by \citetalias{NSS}.

Figure~\ref{fig:MLR_curve} shows that the limit used by \citetalias{NSS} often underestimates the limiting AMRF values separating between class-II and -III binaries. Therefore, we have adopted a more conservative classification curve based on the upper envelope of an ensemble of models generated over various stellar ages and metallicities. As opposed to the \citetalias{NSS} classification curve, our curve provides reliable $\mathcal{A}_{\textsc{ms}}$ and $\mathcal{A}_{\textsc{tr}}$ curves that can be used regardless of the underlying age and metallicity of the binary. We elaborate on the derivation of the curves in the following subsection.

\subsection{Adopted classification limits}
\label{sec: adopted limits}

In Section~\ref{sec: age metallicity based limits} we show that the shape of $\mathcal{A}_{\textsc{ms}}$ and $\mathcal{A}_{\textsc{tr}}$ depends on the age and composition of the stars in the binary system. In light of this claim, a plausible course of action would be to classify each binary while considering its particular age, iron abundance, and corresponding uncertainty estimates.

However, the \texttt{Flame} stellar ages tend to have large uncertainties \citep{creevey22,babusiaux22} and the \texttt{GSP-phot} metallicities are probably biased and require further calibration \citep{andrae22}. Furthermore, these values are provided by \gaia only to about half of the sample of astrometric binaries. As a result, and after attempting to incorporate these values, we concluded that the use of individual age and metallicity estimates is not efficient. Instead, we opted to derive a `global' limiting curve that can provide a conservative estimate for the values for the classification, even when the age and composition are not well constrained.

To do so, we generated a set of $\mathcal{A}_{\textsc{ms}}$ and $\mathcal{A}_{\textsc{tr}}$ curves, spanning from $7$ to $10.2$ in $\log({\rm Age}/{\rm yr})$ and $-3$ to $0.5$ in [Fe/H]. The spacing in both grids is $0.05$ dex. Based on this set of limiting values, we generated a new limiting curve that follows the outer envelope of all curves in our grid. To do so, we used the $99.9$ percentile of all curves for a given mass value and smoothed the resulting envelope with a moving average with a width of ${\sim}0.1$ $\solarmass$. 

Theoretical models are known to estimate the radii of M-dwarfs inaccurately \citep[e.g.,][]{morrell19}. Therefore, for primary stars less massive than ${0.5}$ $\solarmass$, instead of using the upper envelope of the theoretical models, we used the one based on the \citet{mamajek13} MLR. We added a positive constant to this low-mass limiting curve to ensure that the final curve is continuous. The resulting limiting curve is plotted, as solid black lines, along with all the models used in our ensemble, in Figure~\ref{fig:Atr curves}. \red{A lookup table with the values of the limiting curves is provided in the supplementary material}.

\red{Figure~\ref{fig:Atr curves} demonstrates how the limiting curves computed based on the \citet{mamajek13} MLR follow the general trend of those generated using MIST isochrones. The figure also shows that some MIST-based curves significantly deviate from this trend. For primary stars more massive than ${\sim} 1$ $\solarmass$, these deviations are mostly caused by mildly evolved stars, that are on the verge of leaving the MS. On the other hand, for primaries less massive than ${\sim}0.5$ $\solarmass$ the deviations mostly represent young stars of high metallicity.}

\begin{figure}
    \centering
	\includegraphics[width=1\columnwidth]{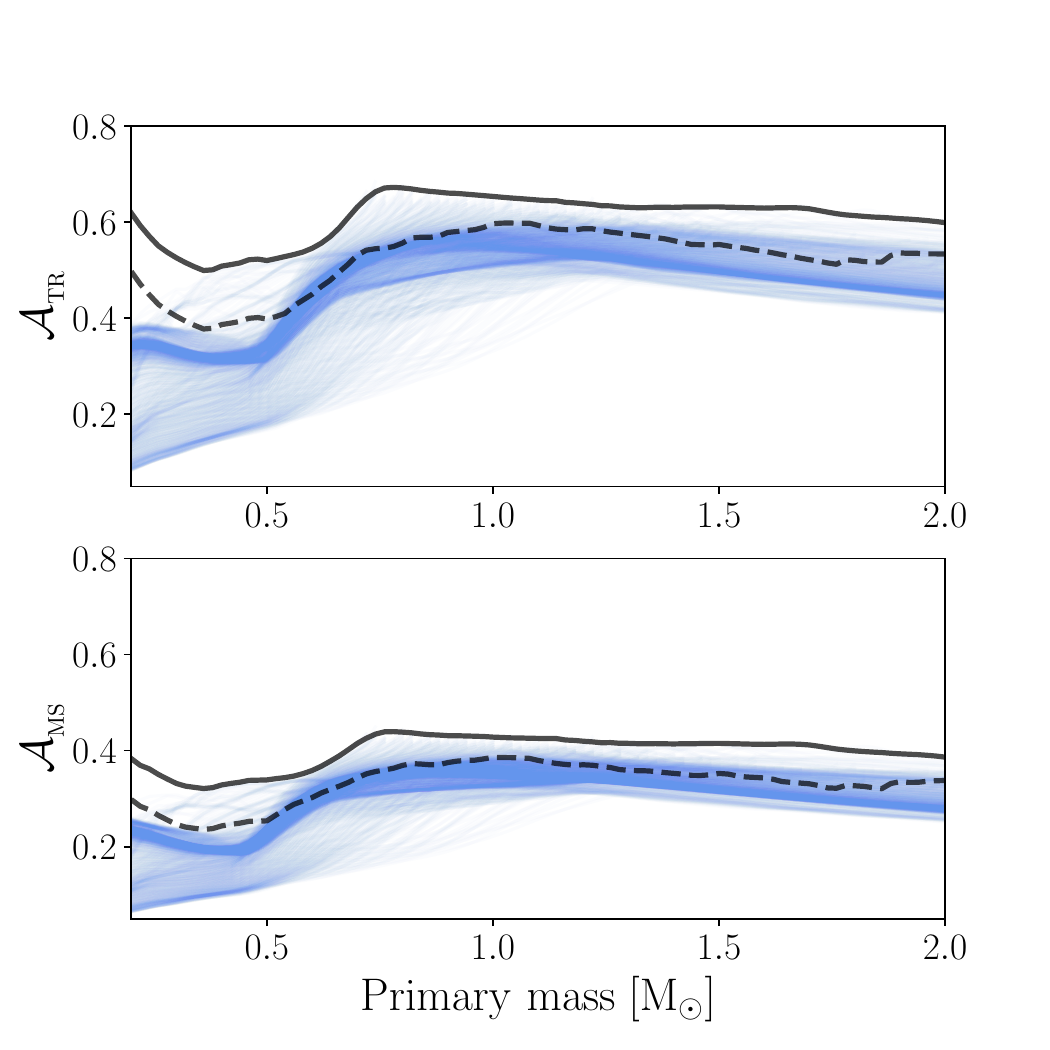}
    \caption{\textit{Top panel}: The limiting AMRF value, $\mathcal{A}_{\textsc{tr}}$ versus the mass of the primary star. An ensemble of models generated from the grid of MIST isochrones appears in light blue. The `global' limit we adopted is plotted as a solid black line. For reference, we also show the limiting curve used by the \citetalias{NSS} team as a dashed black line. \textit{Bottom panel}: Same as the top panel, but for the $\mathcal{A}_{\textsc{ms}}$ curve. }
   \label{fig:Atr curves}
\end{figure}

\section{Triage of \textit{Gaia} binaries}
\label{sec: triage sample}
Equipped with a more conservative threshold for class-III binaries, we now turn to re-consider the \textit{Gaia} astrometric binaries. First, we derive a slightly smaller sample of astrometric binaries by vetting the targets based on the reported orbital parameters. Then, we obtain the probability of each binary being in class-II or class-III, given their orbital parameters and uncertainties.

\subsection{Sample selection}
\label{sec: sample selection}

We first queried the \textit{Gaia} database for astrometric binaries with MS primary stars that have mass estimate, according to the following conditions:
\begin{enumerate}
    \item \texttt{nss solution type} is \texttt{Orbital} or \texttt{AstroSpectroSB1}; \vspace{0.12cm}
    \item \texttt{bit index} is $8191$ or $65535$; \vspace{0.12cm}
    \item \texttt{binary masses} catalogue \texttt{m1 ref} is \texttt{IsocLum}; and \vspace{0.12cm}
    \item \texttt{binary masses} catalogue \texttt{combination method} is
    \item[] \hspace{11pt} \texttt{Orbital+M1} or \texttt{AstroSpectroSB1+M1}.
\end{enumerate}

The first two conditions require that the astrometric orbit was derived from the primary processing pipeline and has all orbital parameters fitted. The following two conditions require that a primary mass estimate exists for the system and that its primary star was classified by as an MS star. \red{For details regarding the derivation of the masses, see section 5.1 of  \citetalias{NSS}}. This procedure left a total of $\var{N_initial}$ targets in the sample.

We then applied the \citet{halbwachs22} criteria on the eccentricity error, parallax \texttt{significance}, $\varpi/{\Delta \varpi}$, and astrometric solution \texttt{significance}, $\alpha_0/{\Delta \alpha_0}$. 
\begin{enumerate}
    \item  $\Delta e  < 0.079\ln{(P/ {\rm day})}-0.244$; \vspace{0.12cm}
   \item $\varpi/{\Delta \varpi} > 20000\cdot(P/{\rm day})^{-1}$; and \vspace{0.12cm}
    \item $\alpha_0/{\Delta \alpha_0} > 158 \cdot (P/{\rm day})^{-1/2}$.
\end{enumerate}
These additional cuts, which were supposed to reduce the number of spurious orbital solutions in the sample, removed only a few additional systems, and we were left with $\var{N_Halbwachs_filter}$ stars with mass estimates. \red{Out of this sample,  $19712$ orbits were obtained based on joint-modelling of the astrometric and spectroscopic data (\texttt{AstroSpectroSB1}) and the rest considered the astrometric data alone (\texttt{Orbital}).
}

Next, we opted to exclude systems with poorly constrained Thiele-Innes coefficients. A full description of these coefficients and the required formulae for using them to derive the angular semi-major axis can be found in \citet{halbwachs22}. We required, somewhat arbitrarily, that the quadratic mean of the relative uncertainty in $A$, $B$, $F$, and $G$ Thiele-Innes coefficients will be smaller than $3$, namely
\begin{equation}
\label{eq: TI_selection_critera}
     \sigma_{\textsc{ti}}^2 \equiv\bigg( \frac{\Delta A}{A} \bigg)^2 + \bigg( \frac{\Delta B}{B} \bigg)^2 + \bigg( \frac{\Delta F}{F} \bigg)^2 + \bigg( \frac{\Delta G}{G} \bigg)^2 \leqslant 36.
\end{equation}
This step was taken to ensure that our classification probability estimates (see below) properly converge and left $\var{N_TI_filter}$ orbits in the cleaned sample.

Finally, we opted to exclude targets with orbital periods longer than the time span of the data analyzed by \gaia DR3. We, therefore, removed systems with orbital periods longer than $1000$ days. We were eventually left with $\var{N_period_filter}$ systems in our cleaned sample.
\red{This sample includes  $16609$ \texttt{AstroSpectroSB1} orbital and $84771$ \texttt{Orbital} solutions. Our additional selection criteria, therefore, slightly favor \texttt{AstroSpectroSB1} solutions over \texttt{Orbital} ones. 
}

\subsection{\red{Distribution of the derived AMRF}}

\begin{figure}
        \centering
        \includegraphics[width=1\columnwidth]{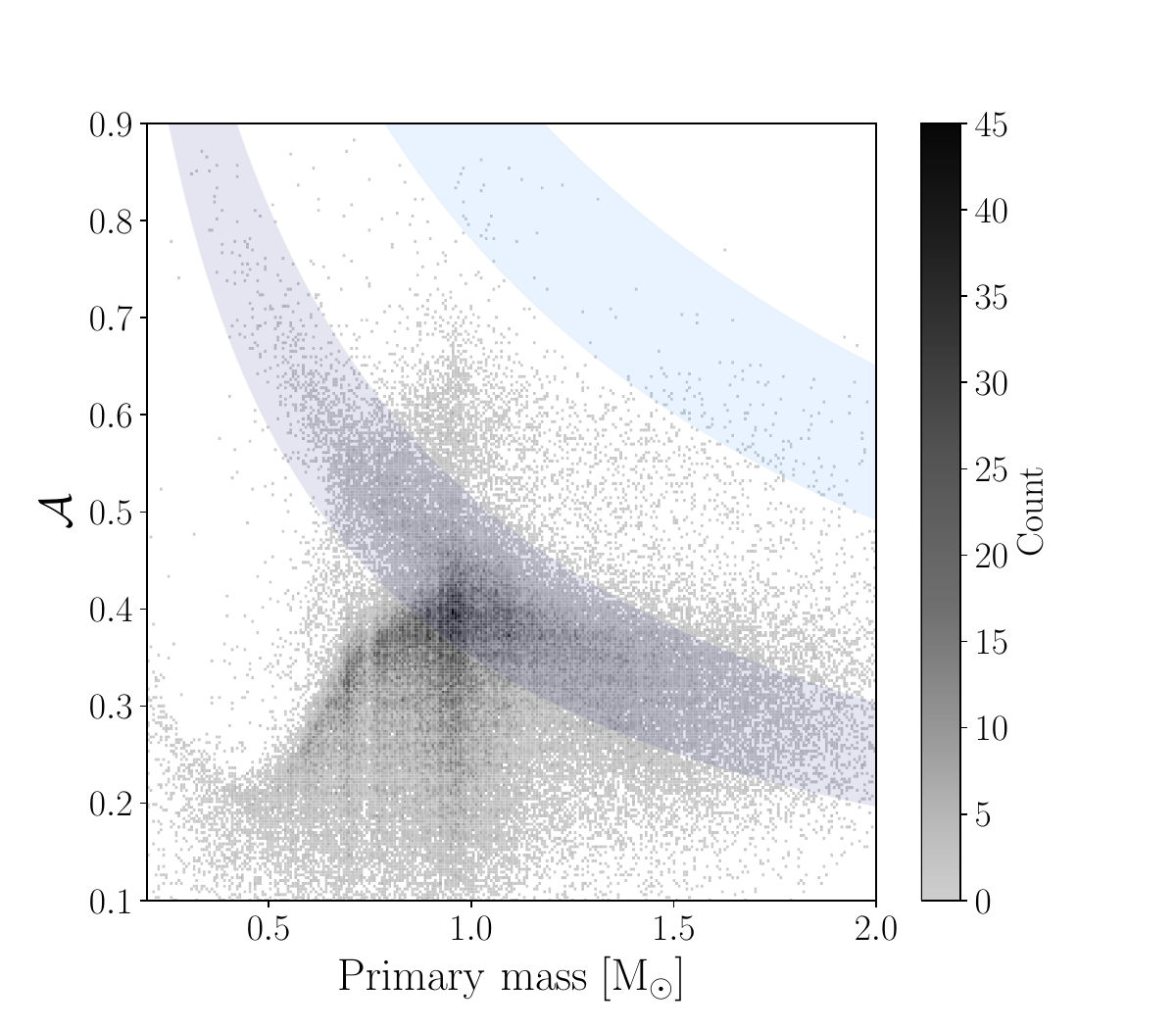} 
\caption{Density plot of the clean astrometric sample in the AMRF--primary-star mass plane, colour coded by the number of points per unit area. Purple and light-blue stripes represent the AMRF locus for typical WDs and NSs (see Figure~\ref{fig:amrf_curve}). Axes ranges were selected to clearly visualize the main locus of the distribution.
}
\label{fig:AMRF_sample}
\end{figure}

Figure~\ref{fig:AMRF_sample} presents a density plot of the cleaned sample on the AMRF--primary-mass plane. 

\red{
The figure displays a prominent vertical concentration at about  ${\sim} 1 \,\, \solarmass$, probably due to the overabundance of solar-type stars in the \gaia sample. The vertical stripe has a clear maximum density at $\mathcal{A} \sim 0.4$, which seems to leak over neighbouring masses, at a range of $\mathcal{A}$   between $0.3$ and $0.4$.
}
\red{
The position and shape of this feature are in line with the expected values of $\mathcal{A}_{\textsc{ms}}$, shown in the bottom panel of Figure~\ref{fig:Atr curves}.
The occurrence of systems at this region of parameters is probably enhanced by an observational bias: for an MS binary, the AMRF attains its maximal value, $\mathcal{A}_\textsc{ms}$ together with the maximal size of the photo-centric orbit. As a result, \textit{Gaia} probably favours the detection of these systems.}

\red{
An additional feature of Figure~\ref{fig:AMRF_sample} is a well-separated cluster of relatively high AMRF values, centered at $\mathcal{A} \sim 0.55$ for primary masses of  ${\sim} 1 \,\, \solarmass$. The position of this cluster is consistent with the expected range of values of $\mathcal{A}_{\textsc{tr}}$, shown in the top panel of Figure~\ref{fig:Atr curves}. A plausible claim is that this cluster is comprised of triple systems with close equal-mass MS binaries as the astrometric secondaries. }

\red{
The diagram also shows an excess of systems with high AMRF values, between $0.45$ and $0.75$ $\solarmass$, located within the purple stripe, 
probably consisting of WD companions, and a few binaries that might have NS companions (see below).
}

\begin{figure}
        \centering
     \includegraphics[trim={0.0cm 2cm 0.0cm 3cm},clip, width=0.49\textwidth]{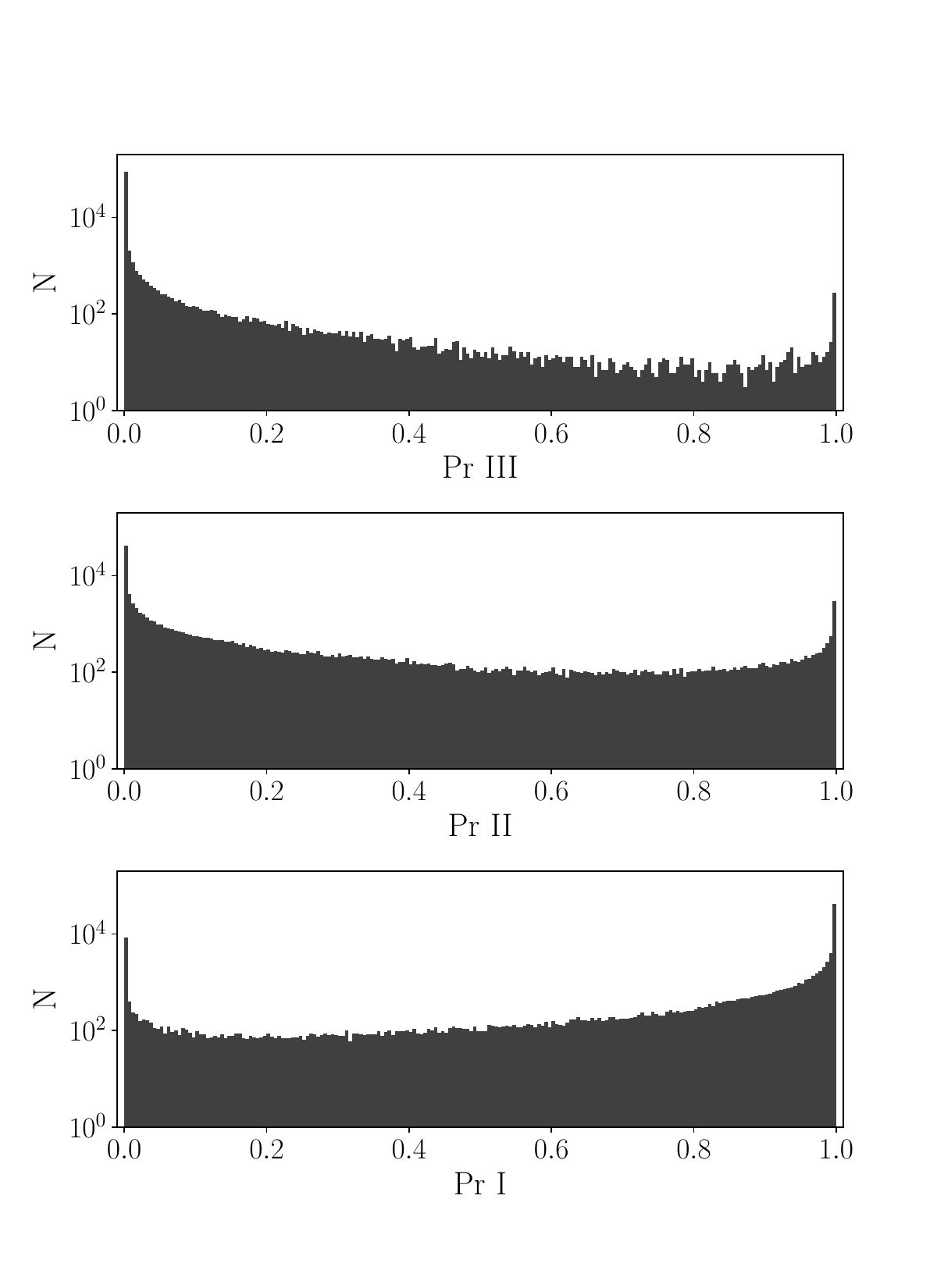}
     \caption{ Histograms of derived  class-III (top panel), class-II (middle panel) and class-I (bottom panel) classification probabilities for the sample of 101380 binaries, reported in Table~\ref{table: prob table}. Bin width in all histograms is $0.5\%$. The left-most and right-most bins of the Pr III histogram contain 87579 and  270 \textit{Gaia} binaries, respectively. The minimal class-III probability in the sample is $99.984\%$. The left-most and right-most bins of the Pr II (Pr I) histogram contain 41578 (8497) and  3009 (41311) \textit{Gaia} binaries, respectively. }
     \label{fig:hist}
    \end{figure}

\begin{figure}
    \centering
	\includegraphics[width=0.495\textwidth]{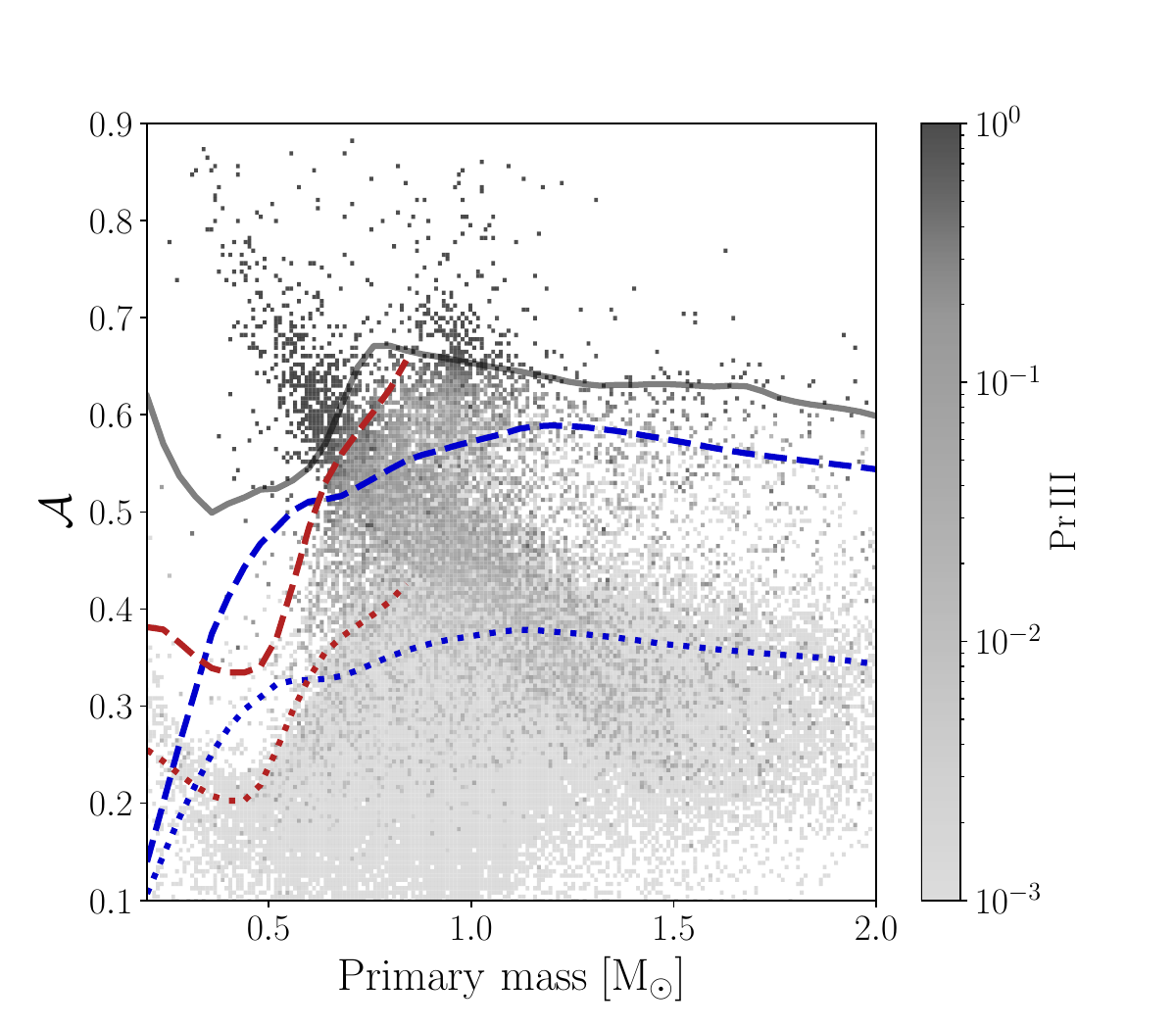}
    \caption{
    Average Pr III (colour coded) on the  $\mathcal{A}$--primary-mass plane for the \textit{Gaia} astrometric sample.  Curves indicate the $\mathcal{A}_{\textsc{ms}}$ and $\mathcal{A}_{\textsc{tr}}$ limits for young (blue lines) and old populations (red lines), as in Figure~\ref{fig:MLR_curve}. The grey line is our selected $\mathcal{A}_{\textsc{tr}}$ curve (see Section~\ref{sec: adopted limits}).}
    \label{fig:MLR_curve2}
\end{figure}

\subsection{Classification probability}
\label{sec: classification prob}

\red{
We move now to dividing the cleaned sample into the three classes of Section~\ref{sec:triage}. Because of the uncertainties of the theoretical boundaries and the uncertainties of the primary mass and $\mathcal{A}$, our classification is of probabilistic nature.
We derive three probabilities:
}

\begin{align}
    {\rm Pr}\,{\rm I} &= {\rm Pr} \big({\mathcal{A} <\mathcal{A}_{\textsc{ms}}} ),  \nonumber \\
     {\rm Pr}\,{\rm II} &\equiv {\rm Pr} \big({\mathcal{A}_{\textsc{ms}} <\mathcal{A} < \mathcal{A}_{\textsc{tr}}} ), \text{ \,\, and} \\
     {\rm Pr}\,{\rm III} &\equiv {\rm Pr} \big({\mathcal{A}_{\textsc{tr}} <\mathcal{A} }),  \nonumber
\end{align}
%
 using Monte-Carlo experiments.
To consider the uncertainties of the orbital elements, we randomly drew $N=10^5$ random instances of the Thiele-Innes parameters, parallax, period, eccentricity and primary mass. The sampling was performed while considering the uncertainties and covariance between the parameters, as reported in the \textit{Gaia} catalogue. 

As proposed by \citet{shahaf19}, the values of the AMRF and primary mass determine the classification of the binary. We calculated $\mathcal{A}$, $\mathcal{A}_{\textsc{ms}}$, and $\mathcal{A}_{\textsc{tr}}$ for each draw, and estimated the class-III probability by
\begin{equation}
    \hat{{\rm Pr}}\,{\rm III}=\frac{r+1}{N+1}\, ,
    \label{eq: pr_bootstrap}
\end{equation}
where $r$ is the number of instances for which $\mathcal{A}$ is larger than $\mathcal{A}_{\textsc{tr}}$ (see, for example, \citealt{davison1997}). The class-II membership probability, $\hat{{\rm Pr}}\,{\rm II}$, was estimated similarly. For brevity, we do not use the `hat' superscript in the following. We emphasize that whenever membership probability is discussed, we refer to our bootstrap-based estimate, derived according to equation~(\ref{eq: pr_bootstrap}), as described above.

A list of the Pr II and Pr III classification probabilities for all targets in our sample is provided in Table~\ref{table: prob table}.  The class-I membership probability, Pr I, can be derived using the two other class probabilities, the number of Monte-Carlo samples, $N$, and equation~(\ref{eq: pr_bootstrap}). Histograms of the classification  probabilities are plotted in Figure~\ref{fig:hist}. 

\red{
Figure~\ref{fig:MLR_curve2} displays the distribution of Pr III in 
the AMRF--primary-mass plane.
Each bin in the diagram is colour coded according to its mean Pr III value.  
A concentration of high-Pr III systems is located above the $\mathcal{A}_\textsc{tr}$ curve, and appears in black. A stripe of systems with high to intermediate class-III probability follows the expected WD envelope. The analysis of this sub-sample of possible binaries with WD companions is deferred to a follow-up study. 
}

Figure~\ref{fig:MLR_curve2} also suggests that for primary stars less massive than ${\sim}0.5$ $\solarmass$, the limiting values calculated according to the MIST models tend to underestimate the transition between class-I and -II systems. This is in accord with a reported discrepancy between the empirically estimated and the theoretically expected M-dwarf radii \citep[e.g.,][]{morrell19}. As described above, we rectified our limiting curves for primaries less massive than $0.5$ $\solarmass$, so that they were not severely affected by this discrepancy (also see Figure~\ref{fig:MLR_curve}).

\begin{table}
\scriptsize
\centering
\begin{tabular}{rrrrrr}
\hline \hline
\multicolumn{1}{c}{Source ID} &    \multicolumn{1}{c}{$M_1$} &   \multicolumn{1}{c}{ $\mathcal{A}$} & \multicolumn{1}{c}{ $M_{2,\min}$} &  \multicolumn{1}{c}{Pr II} & \multicolumn{1}{c}{Pr III} \vspace{0.25pt} \\
\multicolumn{1}{c}{         } &    \multicolumn{1}{c}{($\solarmass$)} &   \multicolumn{1}{c}{    } & \multicolumn{1}{c}{($\solarmass$)} &  \multicolumn{1}{c}{(\%)} & \multicolumn{1}{c}{(\%)} \\
\hline
  33711199137024 &  0.95 &  $0.426(77)$ &  $0.55(13)$ & 56.198 &  8.800 \\
 148953761446272 &  1.25 & $0.3850(70)$ & $0.634(15)$ &  0.836 &  0.001 \\
 301614079110400 &  1.01 &  $0.580(17)$ & $0.895(38)$ & 75.302 & 24.699 \\
 858688517149056 &  1.22 &  $0.196(18)$ & $0.274(28)$ &  0.001 &  0.001 \\
1729398647131392 &  0.56 &  $0.177(23)$ & $0.112(17)$ &  0.072 &  0.001 \\
2488955023504768 &  0.49 &  $0.145(18)$ & $0.078(11)$ &  0.001 &  0.001 \\
3019435024120576 &  0.65 &  $0.299(37)$ & $0.238(35)$ & 36.036 &  2.885 \\
3205080690546176 &  0.85 &  $0.397(34)$ & $0.445(49)$ & 19.105 &  0.006 \\
3334754343120640 &  0.64 &  $0.704(78)$ &  $0.75(13)$ &  6.864 & 93.137 \\
3616877859431808 &  0.93 &  $0.397(17)$ & $0.492(28)$ & 10.752 &  0.001 \\
\hline \hline
\end{tabular}
\caption{Probabilistic AMRF classification of the clean astrometric sample. The G-band magnitude is taken from the \textit{Gaia} DR3 \texttt{source}, and the primary mass is from the \texttt{binary masses} table. The AMRF and its corresponding minimal secondary mass are provided with their uncertainty estimate. The last two columns represent the class-II and -III probability, calculated according to Section~\ref{sec: classification prob}. The full table is available in the online supplementary data.}
\label{table: prob table}
\end{table}

 \begin{table*}
\centering
\begin{tabular}{rrcrrrrll}
\hline \hline
\multicolumn{1}{c}{Source ID} & \multicolumn{1}{c}{$M_2$} &  \multicolumn{1}{c}{$M_1$} & \multicolumn{1}{c}{Period} & \multicolumn{1}{c}{Eccentricity} & \multicolumn{1}{c}{$s$} & \multicolumn{1}{c}{$\sigma_{\textsc{ti}}$ } & \multicolumn{1}{c}{label} & \multicolumn{1}{l}{note}\vspace{0.15pt}\\
\multicolumn{1}{c}{ } & \multicolumn{1}{c}{($\solarmass$)} &  \multicolumn{1}{c}{($\solarmass$)} & \multicolumn{1}{c}{(day)} & \multicolumn{1}{c}{  } & \multicolumn{1}{c}{ } & \multicolumn{1}{c}{    } & &\\
\hline
4373465352415301632 & $12.8(2.6)$ & 1.0 &  $185.77(31)$ &  $0.489(74)$ &         13.6 &    1.2 &    BH &        \textit{Gaia} BH1 \citep{el-badry22c}\\
6281177228434199296 & $11.9(1.5)$ & 1.0 &  $153.95(36)$ &  $0.180(42)$ &         24.3 &    0.6 &    BH &         refuted \citep{el-badry22c}\\
3509370326763016704 &  $3.69(24)$ & 0.7 & $109.392(65)$ &  $0.237(16)$ &         76.1 &    0.2 &    BH &         refuted \citep{el-badry22c}\\
6802561484797464832 &  $3.08(84)$ & 1.2 &  $574.8(6.2)$ &  $0.830(71)$ &          6.8 &    0.3 &    BH &         refuted \citep{el-badry22c}\\
3263804373319076480 &  $2.75(50)$ & 1.0 &  $510.7(4.7)$ &  $0.278(23)$ &         18.1 &    2.1 &    BH & \texttt{AstroSpectroSB1} \\
6601396177408279040 &  $2.57(50)$ & 1.0 &  $533.5(2.0)$ &  $0.791(43)$ &         10.8 &    1.1 &    BH &                 \\
6328149636482597888 &  $2.45(20)$ & 1.1 &     $736(12)$ &  $0.135(36)$ &         89.9 &    3.8 &    BH &                 \\
6588211521163024640 &  $2.41(40)$ & 1.1 &     $943(45)$ &   $0.97(12)$ &         10.4 &    4.2 &    BH &                 \\
4482912934572480384 &  $1.84(19)$ & 0.9 &  $182.39(35)$ &  $0.703(39)$ &         19.3 &    0.7 &    NS &                 \\
5580526947012630912 &  $1.83(25)$ & 1.2 &  $654.3(4.9)$ &  $0.761(40)$ &         12.9 &    0.1 &    NS &                 \\
\hline \hline
\end{tabular}
    \caption{A table of the highly probable class-III systems. The secondary mass of the compact object candidates, derived from the AMRF, appears with its uncertainty. The primary mass, orbital period, eccentricity and significance (denoted $s$) are taken from the \gaia archive. The quadratic mean of the relative error on the Thiele-Innes coefficients denoted $\sigma_{\textsc{ti}}$. The second column from the right presents the result of our tentative Gaussian-mixture based classification. The full table is available in the online supplementary data.
    }
    \label{tab:co table}
\end{table*}

\section{Highly-probable class-III systems}
\label{sec: c3}

We now define a sample of systems likely to host compact companions, applying the \citet{benjamini95} false-discovery rate (FDR) approach, designed to control the expected proportion of false discoveries. In the present work context, false discoveries are systems that are wrongfully identified as class-III binaries. 

We set $\alpha$, the upper limit on the expectancy values of the false discovery rate, to
\begin{equation}
{\alpha}  = \var{BH_alpha} \%\,,
\end{equation}
which yields $\var{N_class3}$ systems in this sub-sample, which we refer to as the class-III sample henceforth. Accordingly, only $18$  (${\sim}\alpha \times \var{N_class3}$) or fewer binaries are expected to be wrongly identified as class-III systems.  

\red{A discussion of what constitutes a false discovery in the context of this work is given in Section~\ref{sec: discussion}}. The selection criterion we used is equivalent to setting a minimal class-III probability of $99.984\%$, i.e., only $16$ out of the $10^5$ Monte-Carlo instances fell below the $\mathcal{A}_{\textsc{tr}}$ limit. 
A list of the selected class-III binaries is given in Table~\ref{tab:co table}.

Most of these systems, if their orbits are valid, contain compact secondaries.  Therefore, we can derive their masses and possibly distinguish between the WD, NS or BH companions. However, we stress the possibility that erroneous orbital fits might contaminate the sample, particularly when considering a sample of rare candidates. We therefore advocate that the validity of these orbits should be assessed externally (see the caveats discussion in Section~\ref{sec: discussion}).

\red{
The proportion of \texttt{AstroSpectroSB1} orbits in the class-III sample is lower than that of the entire sample; out of the 177 systems, only seven binaries have a joint astrometric and spectroscopic orbital solution. 
}
\red{
This is probably because the median G-band magnitude of the class-III systems is  ${\sim}15.6$, ${\sim}2$ magnitudes fainter than the median of the entire clean-astrometric sample. As we know, the \gaia RVS measurements are limited  to bright stars, with a  limit at $\sim14$ mag.
The difference in apparent magnitude between the clean sample and the class-III sample is associated with the mass bias of the triage scheme. As detailed in Sections~\ref{sec:triage} and~\ref{sec: discussion}, the triage is more sensitive to low-mass stars with WD companions. 
}

\subsection{Comparison with the NSS candidates}

\textit{Gaia} DR3 \citetalias{NSS} includes a list of $735$ class-III systems, while our list includes only $177$ binaries. The difference emanates from:

\begin{enumerate}
    \item vetting the quality of the orbital solution,
    \item conservatively  estimating the limiting $\mathcal{A}_{\textsc{tr}}$ curve, and
    \item setting a high-purity threshold on Pr III.
\end{enumerate}

As a result of the different vetting, only $\var{c3_nss_clean}$ systems of the \citetalias{NSS} sample are included in our cleaned sample (see Section~\ref{sec: sample selection}). While all these systems have ${\rm Pr}\,{\rm I}  \simeq 0$, only $\var{c3_nss_and_now}$ were classified here as highly probable class-III systems. We attribute the difference to our conservative approach in setting the limiting classification value, $\mathcal{A}_{\textsc{tr}}$. 

\red{
There are $\var{c3_not_nss_and_now}$ systems in our class-III sample that do not appear in the \citetalias{NSS} class-III sample.
These systems were not included by \citetalias{NSS} because their \texttt{significance} value is smaller than $20$, the limit they adopted for considering valid orbits. As explained above, we used a different limit, which we believe is more appropriate for our purpose, and allowed us to include them in the analysis.}

\subsection{Comparison with the Andrews et al. (2022) candidates}
Another catalogue of $24$ NS and BH candidates in \gaia DR3 was recently published by \citet{andrews22b}, based on their derived mass function (see equation~\ref{eq: fm}); out of this sample, $14$ of are also included in our class-III sample. The remaining $10$ systems were rejected in our early stage of initial sample selection (see Section~\ref{sec: sample selection}) --- 
\red{six have orbital periods longer than $1000$ day, three do not have a primary mass estimate in the \texttt{binary masses} table, and one system did not meet our Thiele-Innes relative uncertainty criterion.}

All $14$ systems shared by both samples appear in our class-III sample. \red{Of these, the companion of Gaia DR3} 6328149636482597888 has a mass larger than  $2.4$ $\solarmass$ and is considered a BH candidate (Table~\ref{tab:co table}). The remaining $13$ are NS candidates, with companion masses between ${\sim}1.2$ and ${\sim}1.8$ $\solarmass$. 

\subsection{Comparison with El-Badry et al. (2022) candidates}

\red{
\citet{el-badry22c} also compiled a list of BH candidates based on \textit{Gaia} DR3 astrometric orbits, using the derived mass ratio of the systems.
Their final list included six targets, including the first two systems of our Table~\ref{tab:co table} class-III sample. Four other systems do not appear on our list, since their orbital periods are longer than $1000$ day.
}

\red{
\citet{el-badry22c} further embarked on an efficient spectroscopic follow-up campaign to validate their orbital solutions.
The first target of Table~\ref{tab:co table}, \textit{Gaia} DR3 4373465352415301632 (\textit{Gaia} BH1 henceforth), was validated by their spectroscopic follow-up campaign. As per the writing of this text,  \textit{Gaia} BH1 is the only bona-fide BH detected in DR3 data. The properties of this system are somewhat unexpected. We refer to \citet{el-badry22c} for a detailed discussion on its properties and the implications of its discovery.
}
\red{
The second target shared by both candidate lists is \textit{Gaia} DR3 6281177228434199296. As opposed to \textit{Gaia} BH1, this system was refuted by the follow-up campaign. }

\red{
\citet{el-badry22c} also monitored, and consequentially refuted, two additional systems identified in our work: \textit{Gaia} DR3 3509370326763016704 and 6802561484797464832. Hence, out of the first four BH candidates presented in Table~\ref{tab:co table}, one was confirmed and three can be deemed as spurious, based on follow-up observations. See the caveats discussion in Section~\ref{sec: discussion}.
}

\section{A tentative distinction between WD and NS candidates}
\label{sec: m2-e}

\begin{figure*}
    \centering
    \begin{minipage}{0.475\textwidth}
        \centering        
        \includegraphics[width=1\columnwidth]{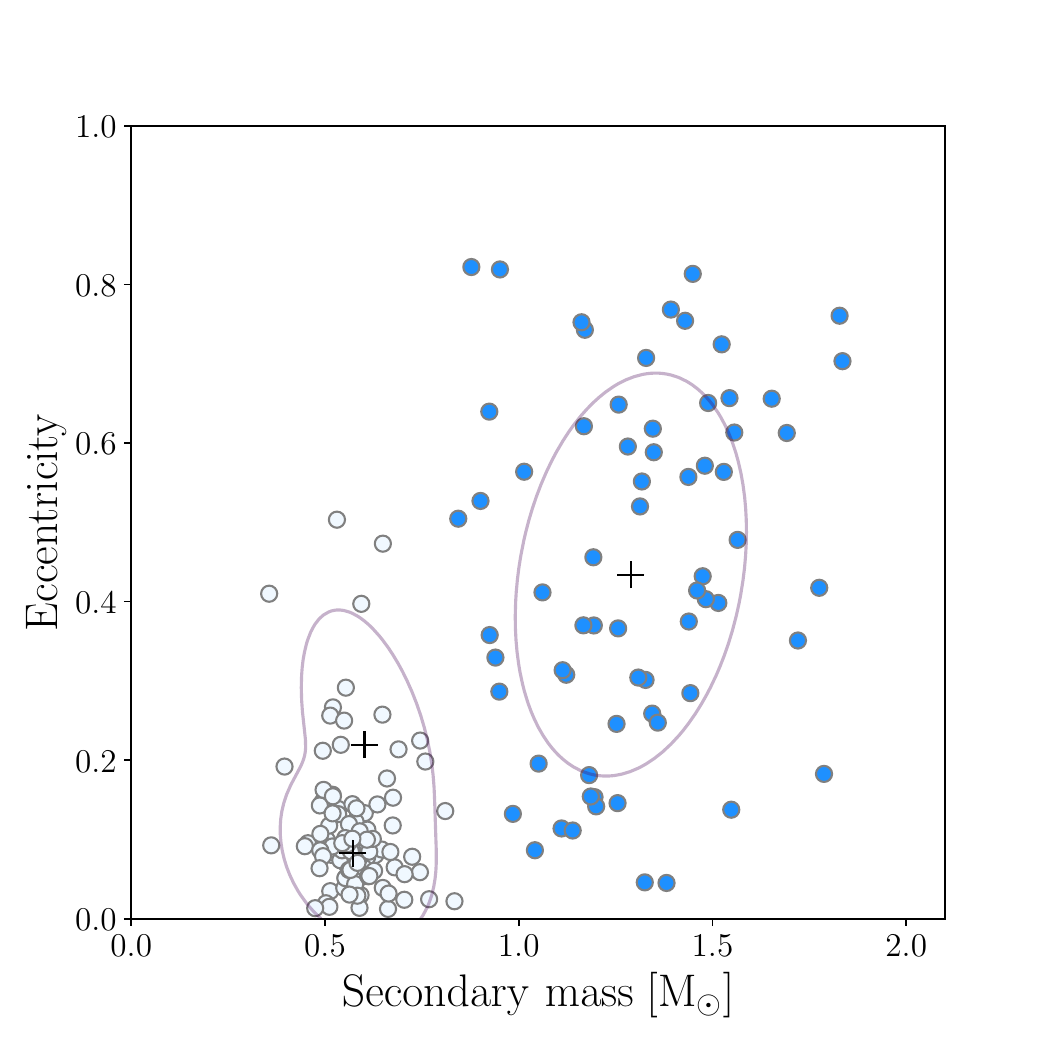} 
        \caption{Orbital eccentricity versus secondary mass for the class-III sample.  We set the upper boundary of the mass axis at $2.1$ $\solarmass$, for clarity. Lines represent the central regions of the fitted Gaussian mixture model (see text). Systems identified as members of the WD and NS clusters appear as white and blue circles, respectively. Lines follow a constant $\ln$-likelihood level of $-0.5$, based on the fitted Gaussian mixture model, and highlight the main loci of the secondary-mass--eccentricity distribution. Centres of the three Gaussian components are plotted as black "+" signs.\newline}
    \label{fig:co_e-m}
    \end{minipage}\hfill
    \begin{minipage}{0.475\textwidth}
        \centering
        \includegraphics[width=1\columnwidth]{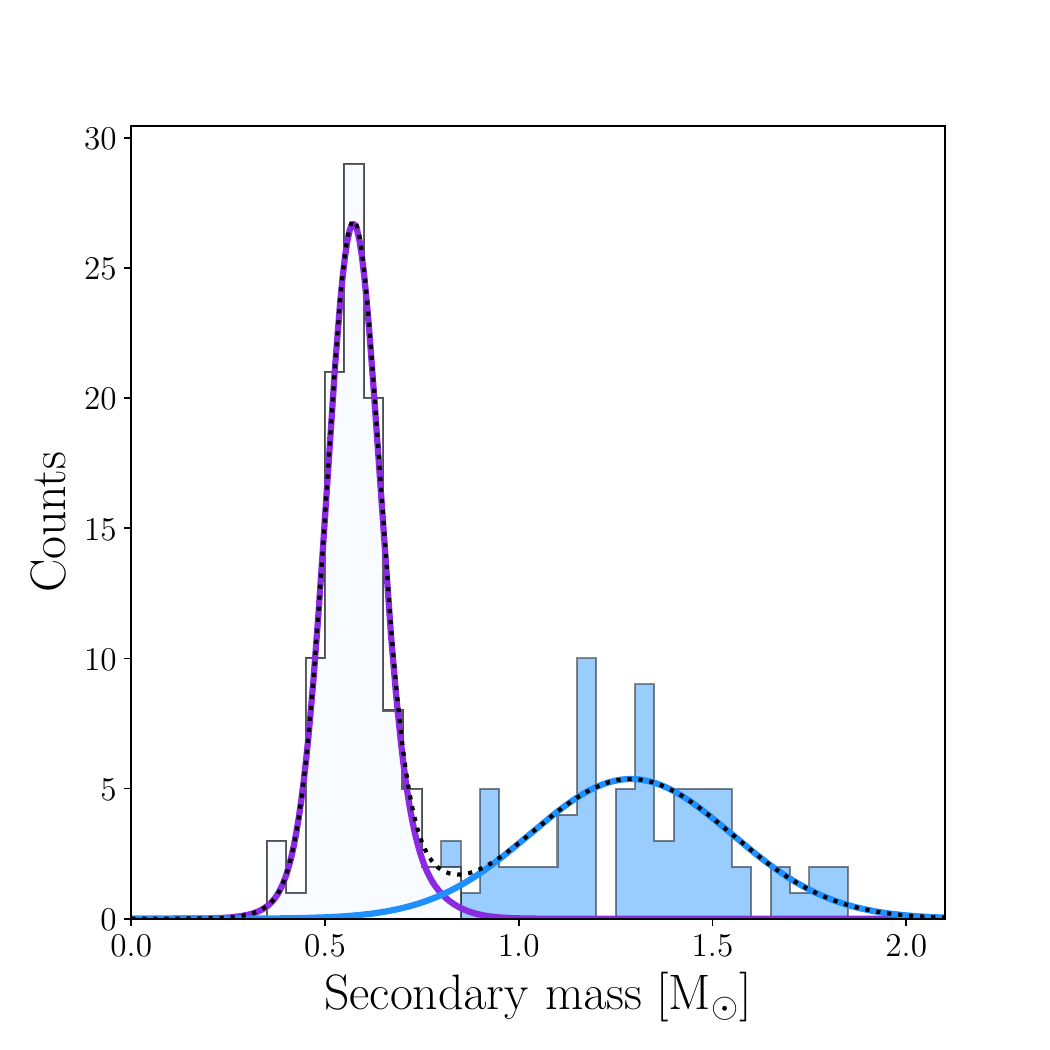} 
        \caption{\red{A stacked secondary-mass histogram of the WD and NS candidates in the class-III sample.} The most prominent peak, at ${\sim}0.6$ $\solarmass$ corresponds to WD secondaries in the sample, and another, less prominent peak, is located at ${\sim}1.3$ $\solarmass$. We set the upper boundary of the mass axis at $2.1$ $\solarmass$, for clarity. As a result, $8$ additional systems with secondary masses larger than $2.4$ $\solarmass$ are not shown. Lines represent the marginal distributions derived from a Gaussian-mixture model that was fitted to the secondary-mass--eccentricity diagram (see Section~\ref{sec: m2-e}). The overlap between the two classes is small, as only one bin shares points from both classes, with two WDs and one NS.
        }
    \label{fig:hist_m2}
    \end{minipage}
\end{figure*}

The distinction between the WD and the NS in our catalogue is not trivial because some WDs were found by previous studies to have masses greater than the masses of the least massive NSs  \citep[e.g.,][]{martinez15,caiazzo21}. Furthermore, some compact secondaries with mass typical of NS might be close binaries composed of two WDs. Nevertheless, one might be helped in separating the two populations if some orbital properties of the WD binaries statistically differ from those of the NS binaries.

To explore this possibility, we plot in Figure~\ref{fig:co_e-m} the orbital eccentricity versus the secondary mass for all objects in our compact-object sample, except for the eight systems with masses larger than $2.4$ $\solarmass$. The figure suggests two clusters of binaries --- one with the typical WD mass of $0.6\,\solarmass$ and low eccentricity, and the other with the typical NS mass of $ 1.2$ $\solarmass$ and higher eccentricity.

To tentatively divide the sample into WD and NS candidates, we fitted the eccentricity--secondary-mass diagram with a component Gaussian mixture model.\footnote{This was done using \texttt{Scikit-learn} \href{https://scikit-learn.org/stable/modules/generated/sklearn.mixture.GaussianMixture.html}{GaussianMixture module}.} 
\red{
The mean Silhouette similarity score \citep{rousseeuw87} was used to select the number of components in the mixture.\footnote{See the \texttt{Scikit-learn} \href{https://scikit-learn.org/stable/modules/generated/sklearn.metrics.silhouette_score.html}{Silhouette score function}.} The score was calculated using the cosine-distance metric, to favour the masses-eccentricities relation over their actual values. The Silhouette score became negative when using more than three Gaussian components, which indicates that the resulting clusters overlap. Nevertheless, we emphasise that this is merely a tentative classification, and refer the readers to the caveats discussion in Section~\ref{sec: discussion}.}

We used two components to describe the distribution of the low-mass circularized systems (`WD cluster') and another component for the massive eccentric ones (`NS cluster').
The WD cluster is described by
\begin{equation}
\begin{aligned}
    f_{\textsc{wd}} \sim 
0.78&\, \mathcal{N}\bigg( 
\begin{bmatrix}
   0.572\\
   0.083
\end{bmatrix}, \begin{bmatrix}
   0.004 & -0.0006\\
 -0.0006 &   0.001
\end{bmatrix}
 \bigg) \, + \\
0.22&\, \mathcal{N}\bigg( 
\begin{bmatrix}
    0.60\\
    0.22
\end{bmatrix}, \begin{bmatrix}
    0.01 & -0.006\\
 -0.006 &    0.02
\end{bmatrix}
 \bigg)\, ,
\end{aligned}
\end{equation}
where $\mathcal{N}$ represents a normal distribution, its first entry representing the derived expectancy for the secondary mass in Solar units (top) and the eccentricity (bottom), and the second entry is the corresponding covariance matrix. Similarly, the NS cluster is described by
\begin{equation}
\label{eq: NS gaussian}
\begin{aligned}
    f_{\textsc{ns}} \sim 
&\, \mathcal{N}\bigg( 
\begin{bmatrix}
    1.29\\
    0.43
\end{bmatrix}, \begin{bmatrix}
    0.07 &    0.01\\
    0.01 &    0.05
\end{bmatrix}
 \bigg)\, .
\end{aligned}
\end{equation}
The odds ratio between the two clusters is $1.42$, in favour of the WD cluster.

The central regions of the two clusters are presented as thin purple lines in Figure~\ref{fig:co_e-m}. The points are coloured by their classification: white circles represent the objects in the WD cluster, and the blue circles represent the objects in the NS cluster. The distinction between the two clusters was made according to their cluster-membership probabilities, 
$f_{\textsc{ns}}$ and $f_{\textsc{ns}}$,
of the Gaussian mixture model. 
The $68$ targets with $f_{\textsc{ns}}>0.5$ were labeled as NS cluster members; the $101$ targets in the complement set, with $f_{\textsc{wd}}>0.5$, were labeled as members of the WD cluster. The $8$ BH candidates, with masses larger than $2.4$ $\solarmass$, are not included in any of the two classes.

Figure~\ref{fig:hist_m2} presents the mass distribution of the secondary masses, overlaid with 
the marginal probability density function of the Gaussian mixture of Figure~\ref{fig:co_e-m}. The solid purple and blue curves represent the distribution of the WD and NS clusters, respectively, and the combined marginal distribution of the entire sample is shown as a black dotted line.

Out of $\var{N_class3}$ binaries in the sample, $\var{n_wd_stripe}$ have companions in the mass range of $0.45{-}0.75$ $\solarmass$. These systems populate a prominent and narrow histogram peak, centred at ${\sim}0.6$ $\solarmass$, which is qualitatively consistent with the observed WD mass distribution \citep[e.g.,][]{tremblay16, hollands18}. The histogram also shows a broad secondary peak, centred at ${\sim}1.3$ $\solarmass$, probably composed on NS secondaries. Note, however, that the high-mass wing of the secondary peak contains $\var{n_ns_stripe}$ systems with companions of $1.4{-}2.1$ $\solarmass$, which 
could also be close binaries by themselves composed of two WDs; see the discussion in \citealt{Mazeh22}). Additional $\var{n_interm_stripe}$ systems populate the intermediate mass range of $0.75{-}1.4$ $\solarmass$, which could either be NSs or massive WDs.

\begin{figure*}
    \centering
    \begin{minipage}{0.475\textwidth}
        \centering        
        \includegraphics[width=1\textwidth]{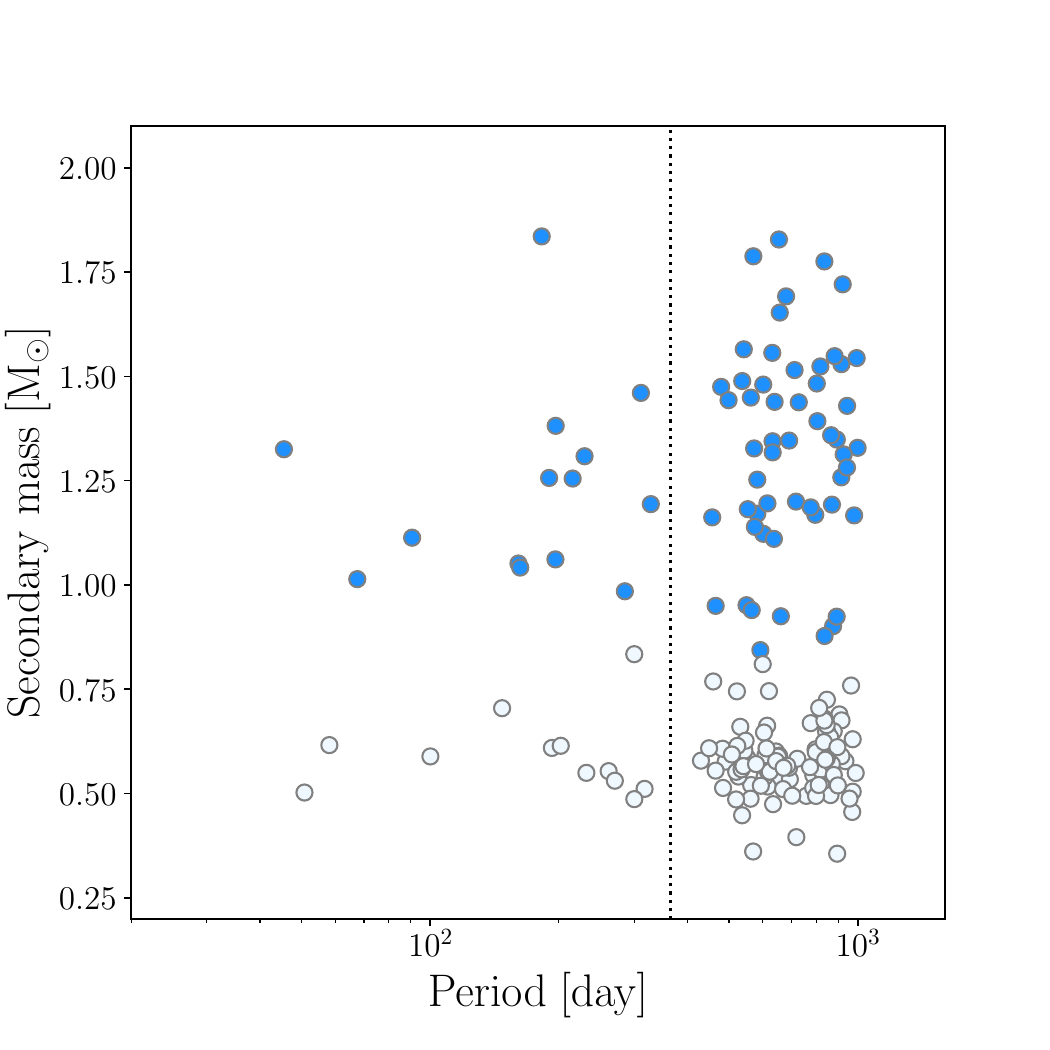} 
        \caption{Secondary mass versus orbital period for the class-III sample. Points are coloured by the classification in Figure~\ref{fig:co_e-m}.  We set the upper boundary of the mass axis at $2.1$ $\solarmass$, for clarity. As a result, $8$ additional systems with secondary masses larger than $2.4$ $\solarmass$ are not shown in the figure. \red{The vertical dotted line corresponds to an orbital period of one year.}}
    \label{fig:co_Pm}
    \end{minipage}\hfill
    \begin{minipage}{0.475\textwidth}
        \centering
        \includegraphics[width=1\textwidth]{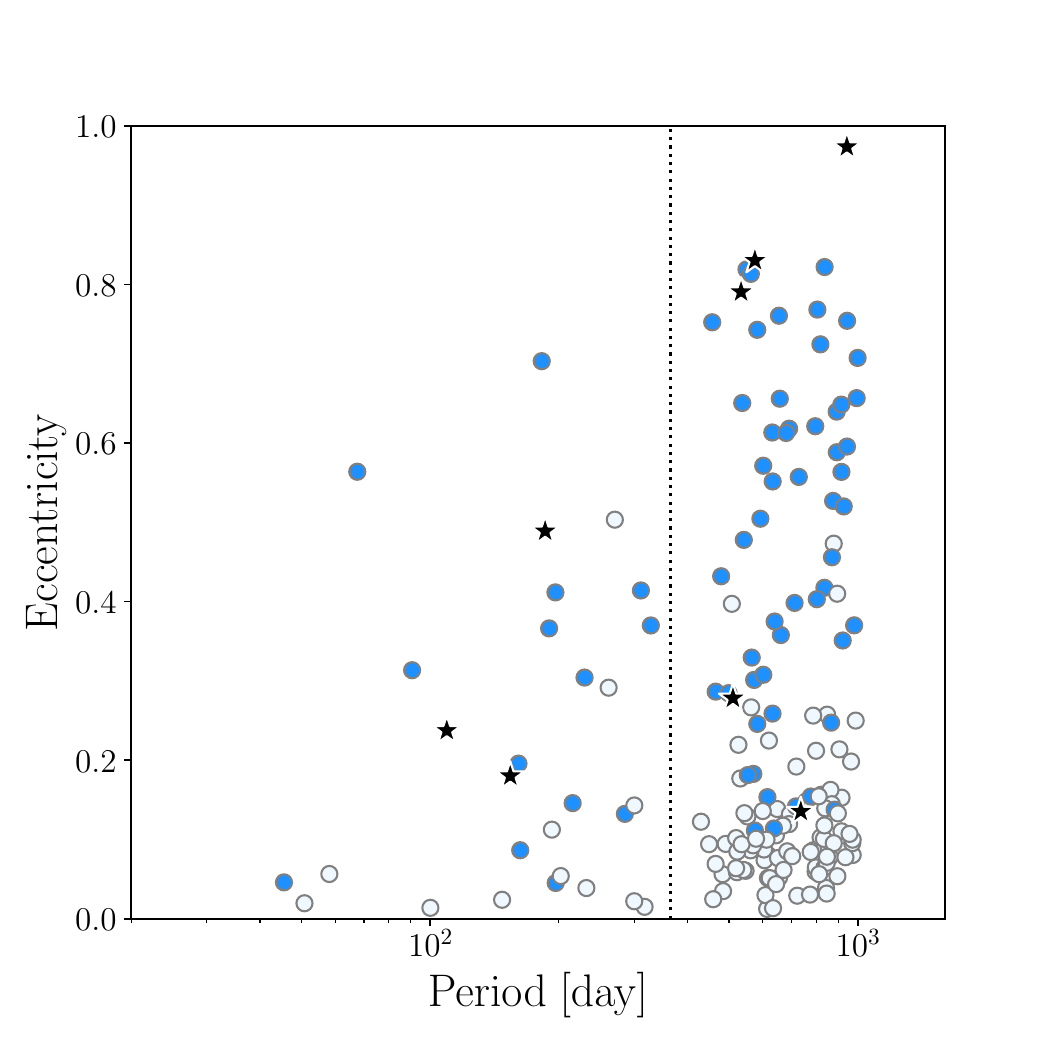} 
        \caption{Period-eccentricity diagram of the class-III sample. Points are coloured by the classification in Figure~\ref{fig:co_e-m}. Eight systems, with secondary masses larger than $2.4$ $\solarmass$, are shown as black stars.  The vertical dotted line corresponds to an orbital period of one year.\newline}
    \label{fig:co_Pe}
    \end{minipage}\hfill \\
\end{figure*}

Figures~\ref{fig:co_Pm} and~\ref{fig:co_Pe} present the secondary mass and eccentricity versus their orbital period, respectively, for the highly-probable class-III systems in our sample. The points in the figures are coloured according to the tentative mass-eccentricity clustering described above. Figure~\ref{fig:co_Pm} suggests that (except for two NS cases, Gaia DR3 4482912934572480384 and 1522897482203494784), the NS candidates are confined below some upper envelope in the mass-period diagram. Similarly, Figure~\ref{fig:co_Pe} suggests that an upper envelope also exists in the period-eccentricity plane (except for Gaia DR3 4482912934572480384 and 2574867704662509568). The sample indicates that the compact object's mass and orbital eccentricity can reach larger values as the orbital period lasts longer.

\begin{figure}
        \centering        
        \includegraphics[width=1\columnwidth]{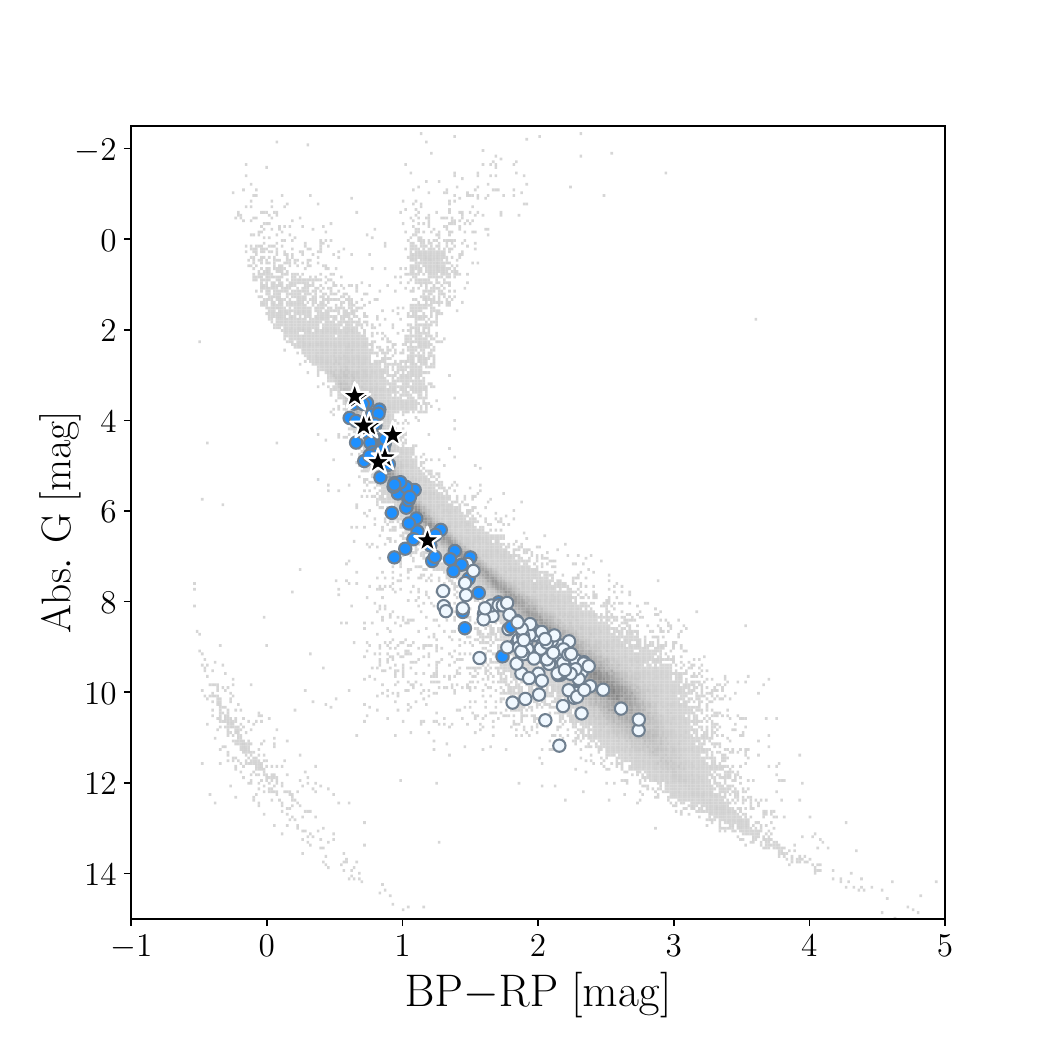} 
        \caption{A colour-magnitude diagram showing the \gaia G-band absolute magnitude versus the BP-RP colour index for the class-III sample. \red{The position of the class-III systems on the diagram accounts for interstellar extinction and reddening, whenever these data were available in DR3.} Points are coloured by their classification in Figure~\ref{fig:co_e-m}. Eight systems with masses larger than $2.4$ $\solarmass$ are shown as black stars. The grey background shows the \gaia Catalogue of Nearby Stars for reference.  }
    \label{fig:co_cmd}
\end{figure}

The CMD location of the compact-object binaries are presented in Figure~\ref{fig:co_cmd}. The figure also shows, for reference, the CMD of the \gaia Catalogue of Nearby Stars \citep[GCNS; ][]{smart21} for all systems brighter than $15$ in \textit{Gaia}'s G band. One can see that all the binaries have MS primaries, as required by our analysis. As a rule, the binaries occupy the bluer part of the MS stripe, and some less massive WD binaries are even slightly bluer than the edge of the neighbouring MS stars. This might be due to some short-wavelength contribution from the WD companions \citep[see, for example, ][]{eyer19}. 

\red{
Interestingly, some NS cluster members are also located on the blue side of the MS stripe. One obvious outlier is \textit{Gaia} DR3 2469926638416055168, with an absolute magnitude of ${\sim}7$ and colour index of ${\sim}0.95$. This binary is  eccentric ($e{\sim}0.75$), with an orbital period of ${\sim}580$ day. The primary mass is ${\sim}0.5$ $\solarmass$, and the companion is of $1.17\pm0.14$ \solarmass. One possibility is that the companion is a massive WD or a close binary composed of two WDs, that was wrongly classified as an NS in our naive Gaussian-mixture classification, due to the high eccentricity of the wide orbit. 
Alternatively, this discrepancy could originate from an incorrect estimate of parallax or interstellar extinction.
}

\subsection{Incompleteness of the compact object sample}

\begin{figure}
        \centering
        \includegraphics[width=1\columnwidth]{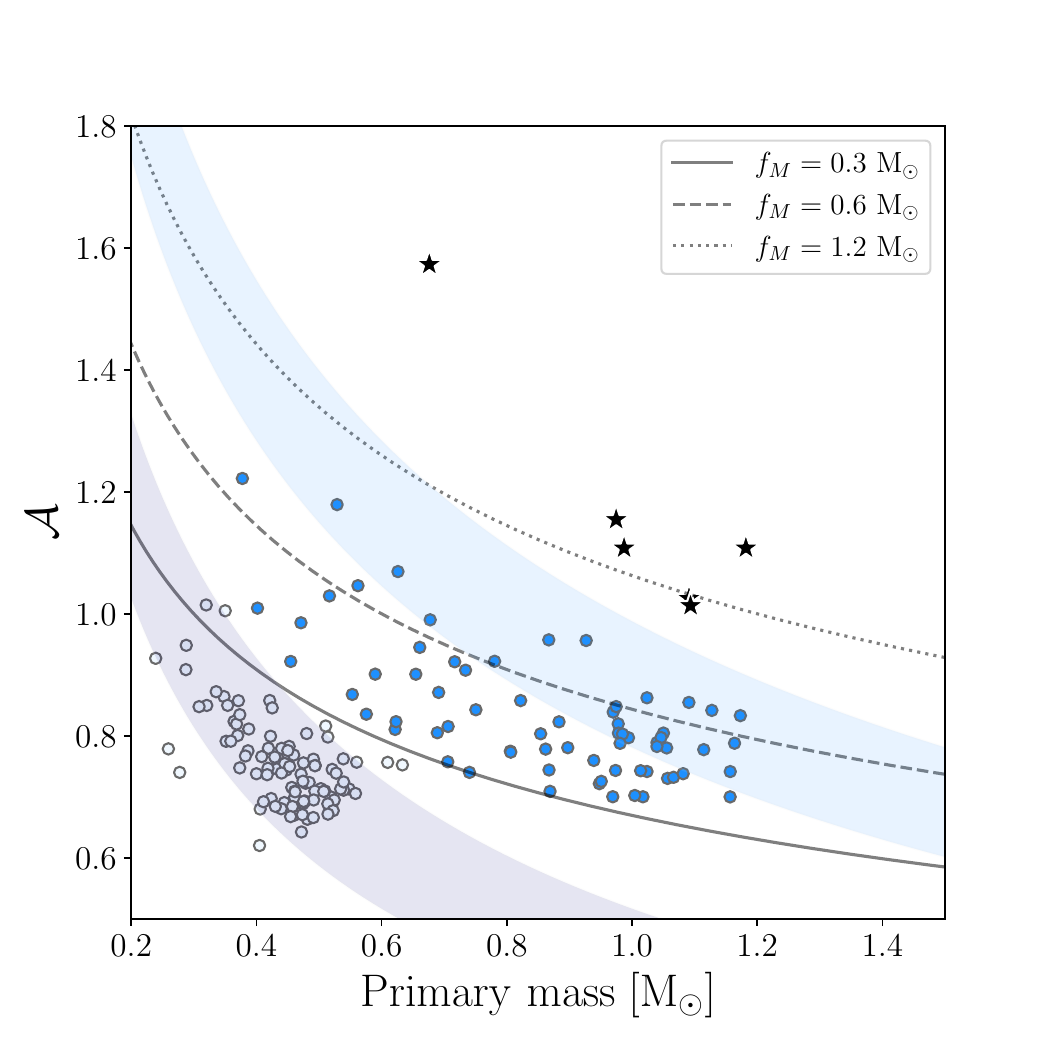} 
        \caption{AMRF as a function of primary mass for the class-III sample. Two BH candidates with AMRF values larger than $1.8$ are not shown. The remaining six systems with masses larger than $2.4$ $\solarmass$ appear as black stars. Purple and light-blue stripes represent the locus of typical WDs and NSs (see Figures~\ref{fig:amrf_curve} and~\ref{fig:AMRF_sample}). Contours of constant mass function, $f_M$, are plotted as solid, dashed and dotted black lines. Note that axes ranges differ from those of previous Figures.}
    \label{fig:III A_m1}
\end{figure}

Figure~\ref{fig:III A_m1} shows the location of the selected binaries on the AMRF--primary-mass plane and illustrates some of the selection biases that affect this sample. The figure shows that all MS-WD binaries, except two cases, have primaries less massive than ${\sim}\,0.6$ $\solarmass$. On the other hand, it seems that the MS-NS binaries tend to have primaries more massive than ${\sim}0.7$ $\solarmass$. This emerging relationship between the mass of the primary and that of the secondary is probably induced by the triage selection scheme: companions in the WD mass range can be identified as class-III binaries only if the mass of their primary host is sufficiently low (also see Figure~\ref{fig:amrf_curve}). 

The sample also presents a significant paucity of BH compared to the recent theoretical predictions \citep[e.g.,][]{mashian17}; only $\var{n_bh_stripe}$ candidates in the class-III sample have companions more massive than $2.4$ $\solarmass$, and a few of them were already refuted. Supposedly, very massive non-luminous companions should have been easily detected by \gaia. However, as \citet{halbwachs22} showed, various properties of \textit{Gaia}'s  orbit and sampling yielded spurious orbital solutions, which were often characterized by high mass functions,
\begin{equation}
    \label{eq: fm}
    f_M = \mathcal{A}^3 M_1 > 0.3 \,\, \solarmass.
\end{equation}
While \citet{halbwachs22} did not explicitly reject systems based on the value of their mass function, it is plausible that many of the BHs and NSs initially detected by \gaia were indistinguishable from spurious solutions and consequentially excluded from \textit{Gaia} binary-star database. 

One possible way of explaining such a bias is by considering the correlation of the parallax error, $\Delta \varpi$, with the photo-centric semi-major axis, $\alpha_0$. Consequently, the selection imposed on the parallax significance and the orbital period (see Section~\ref{sec: sample selection}) can implicitly impose a selection effect on the total mass of the system. To illustrate this point, we overlaid Figure~\ref{fig:III A_m1} with three equal-$f_M$ contours. The occurrence rate of the class-III systems appears to be decreasing along the direction perpendicular to these curves, towards high $f_M$ values. While we cannot rule out that this effect is due to the actual underlying occurrence rates, it is also possible that the population of high-$f_M$ companions was significantly depleted in \gaia DR3.

The mass distribution of the compact object candidates in our sample is therefore heavily biased. However, while some BH and NS were probably excluded from DR3 and could only be recovered in future data releases, the case of WDs is different. Many WD binaries probably exist in the \gaia sample but were identified as class-I/II binaries and consequentially eluded detection. We further discuss the identification of WD secondaries in an accompanying paper. 


\section{ Summary and Discussion}
\label{sec: discussion}
We have applied the triage analysis of \citet{shahaf19} to the recently published sample of \gaia astrometric binaries of \citetalias{NSS}. The analysis divides the astrometric binaries into three classes, class-I --- systems with MS secondary, class-II --- binaries that are likely to be triple systems, with a close MS binary as the astrometric secondary, and class-III --- binaries that probably have a compact-object companion. 

The analysis was based on three levels of computation. First, we vetted some of the orbits, based on the relative errors of the Thiele-Innes astrometric parameters, as published by \citetalias{NSS}, and the recommended selection criteria of \citet{halbwachs22}. We also rejected binaries with periods longer than $1000$ days. Our criteria resulted in $\var{N_period_filter}$ binaries. Second, we adopted a new, conservative, $\mathcal{A}_{\textsc{tr}}$ threshold, based on the MIST stellar evolutionary tracks. Finally, we derived the class-II and class-III probability of each binary, taking into account the uncertainties of the Thiele-Innes parameters and the stellar mass. The main product of this analysis is a catalogue of these astrometric binaries with probabilities to be in each of the classes. 

\red{
Based on the classification probabilities, we constructed a small sample of $\var{N_class3}$ astrometric binaries that are likely to have compact companions. For comparison,  \citetalias{NSS} constructed a larger list of 735 binaries. Another catalogue, by  \citet[][]{andrews22b}, contained 24 systems, and a list of $6$ BH candidates (out of which one was dynamically validated) was provided and by \citet{el-badry22c}.}

\red{
Our sample was chosen such that the expected false-discovery rate is below $\var{BH_alpha}\%$, so we can place an upper limit on the expected number of contaminants, assuming all orbital solutions are valid (but see the caveats discussion below).
In the context of this work, contaminants might be hierarchical triples that were falsely identified as class-III systems. }

The requirements we adopted made our sample of binaries with probable compact objects rather small and incomplete. It is therefore too early to use it to draw conclusions regarding the frequency of binaries with dormant compact companions. However, we already can see some statistical features, plotted in Figures~\ref{fig:co_e-m}--\ref{fig:co_Pe}, that seem real and might be of astrophysical interest.

\subsection{WD, NS and BH binaries}

The new sample includes eight systems with compact-object masses larger than $2.4$ \solarmass, probable binaries with BH companions. This classification is somewhat arbitrary, as the borderline between NSs and BHs is not clear. In fact, six of these candidates reside in what was considered a mass gap between the two types of compact objects \citep[e.g.,][]{MassGap12}. However, this gap started to fill up recently by masses measured through gravitational waves \citep[e.g.,][]{MassGap22a,MassGap22b}. 

\red{
Half of the BH candidates identified in this work were followed-up in a spectroscopic campaign by \citet{el-badry22c}. One system, \textit{Gaia} BH1, was validated based on its radial-velocity (RV) modulation. For a detailed discussion regarding the properties and astrophysical implications of of \textit{Gaia} BH1, see \citet{el-badry22c}. The remaining three were identified as spurious solutions (see Table~\ref{tab:co table}). Several illuminating examples of erroneous orbital solutions or misclassification, resulting in false detections of BH-mass companions were also recently discussed by \citet{bashi22} and \citet{elbadry22} in the context of \gaia spectroscopic orbits. 
}

The validity of this small BH-candidate sample, therefore, requires further study. As mentioned in Section~\ref{sec: m2-e}, most BH candidates have GoF values higher than ${\sim}5$. One orbital solution, Gaia DR3 6588211521163024640, has an exceptionally high eccentricity and an orbital period consistent with ${\sim}1000$ days, which raises suspicions regarding its quality. Validating the orbits of the BH candidates using data sources external to \gaia DR3 is crucial (see the caveats discussion below). Testing the validity of the astrometric orbits is beyond the scope of this work. 

The other 169 systems, with companion masses smaller than $1.85$ \solarmass,  are probably mostly WD or NS binaries. We tried to distinguish between the WD and NS by plotting the orbital eccentricity versus the derived compact-object mass. The diagram suggests a clear separation between the WD and the NS binaries. Most of the WD binaries are characterized by small eccentricities of about $0.1$ and masses of $0.6$ \solarmass, while the NS binaries display eccentricities of about $0.4$ and masses of $1.3$ \solarmass. The latter feature might be due to the natal kicks that accompany the NS formation \citep{hansen97, igoshev19}, although their underlying physical mechanism is a matter of ongoing research \citep{atri19, callister21, willcox21,andrews22a}. 

As a population, the detected binaries in the NS cluster carry the potential of probing the margins of the natal kick velocity distribution that is assumed to be associated with the NS formation. With orbital periods of up to a few years and eccentricities below ${\sim}0.8$, these binaries probably represent the products of processes that, while strong enough to induce eccentricity to the orbit, could not disrupt the binary entirely  \citep[e.g.,][]{pfahl02, heuvel07, beniamini16, tauris17}. The clear dependence of the eccentricity and the NS masses can be used as hints for the nature of the last stages of the orbital formation of these binaries.

As opposed to the NS candidates, most binaries in the WD cluster have 
small orbital eccentricity (see Figures~\ref{fig:co_e-m} and~\ref{fig:co_Pe}). 
WDs are expected to be in circularized orbits due to the tidal interaction between the WD progenitors and their MS companion, and therefore our result could be of interest \citep{zahn77, izzard10, swaelmen17, jorissen19}. 
However, it is too early to determine whether those small eccentricities are significant. 
One could claim, for example, that this is not an inherent property of the sample, as it could originate from the fitting procedure or a sample selection (see the caveats discussion below).

\red{
As pointed out above, the most striking feature of Figures~\ref{fig:co_Pm}  is the concentration of the binaries in a period range of about $400$--$1000$ days. One needs to check whether this results from an observational bias, as the longer the period, the larger the semi-major axis is \citep[see, for example, the discussion by][]{penoyre22}. Similarly, the figures suggest that upper envelopes of the NS distributions with larger mass and eccentricity for orbits with longer periods have surfaced. If this is not another result of an observational bias, it could be the result of the natal kicks discussed above, which might produce statistical dependence between the resulting period and the NS mass and orbital eccentricity \citep{hills83,brandt95,kalogera96, dewi05}.
}

The mass distribution of the WD and NS class-III sample is presented in Figure~\ref{fig:hist_m2} (also see figure 36 of \citetalias{NSS}). The advantage of a sample  of astrometric binaries with compact companions is the ability to dynamically derive the secondary mass of each binary, which depends only on the primary mass and orbital elements, provided the secondary is non-luminous. The derived masses of WD, for example,  do not depend on evolutionary tracks nor on spectral analysis \citep[e.g.,][]{bergeron19, torres21, fantin21, heintz22}.

The secondary-mass histogram presents a sharp peak at ${\sim}0.6\,\solarmass$, with a width of $\sim 0.1\,\solarmass$, similar to the peak found in the distribution of the WD in the solar neighbourhood by \citet{tremblay16} and  \citet[][]{hollands18}. We, therefore, can assume that the peak of Figure~\ref{fig:hist_m2} does reflect the masses of a large sample of WD secondaries (see also \citetalias{NSS}). It seems as if the WD distribution reported by \citet[][]{hollands18} is wider than the one of Figure~\ref{fig:hist_m2}. One might wonder if this is because of the more precise determination of the WD companion mass by dynamical techniques. In any case, it is also possible that the proximity of the companion through the last evolutionary phases leading to the production of WDs might modify the mass of the end product \citep[e.g.,][]{toonen14}. 

The mass distribution of Figure~\ref{fig:hist_m2} includes a wide `wing' to the right of the sharp peak, centered around ${\sim}1.3\,\solarmass$, that we identified as NSs, consistent with their expected masses \citep[e.g.,][]{lattimer14,ozel16}.
The mass-eccentricity diagram suggests that 
within the period range probed by \gaia
the WD and NS mass distributions only slightly overlap --- compact companions more massive than  ${\sim}1$ $\solarmass$ are likely to be NS. 
This observation stands in contrast with 
the recent analysis of the Gaia EDR3 catalog \citep{GentileFusillo21} which argues that the mass distribution of WDs in the solar neighborhood has a long tail extending to $1.2$ $\solarmass$ and even higher. However, the latter distribution was derived for single WDs, for which the massive tail might reflect the result of two merging WDs \citep{kilic21, miller22, fleury22a, fleury22b}.
Our tentative separation, on the other hand, is based on the masses of compact companions, for which the binarity did not allow the merging of two WDs in the close proximity of the optical star.  Such a distinction might have obvious implications for classifying non-interacting compact objects \citep[see, for example,][]{Mazeh22}. 

Many binaries with compact secondaries were previously known, either as cataclysmic variables \citep[e.g.,][]{CV76,CV11}, X-ray binaries \citep[e.g.,][]{paul17} or binary pulsars \citep[e.g.,][]{manchester17}. Most of those (except Be/X-ray binaries and a few pulsars) reside in short-period orbits, on the order of hours and days. Most of them were discovered by the luminosity of the compact objects or the accretion disks around them, which are fueled by mass transfer from the optical companion or the rotational energy of the compact object (but see \citealt[][]{shenar22, shenar22b}). The companions sampled here are all dormant, and their identification is based on the astrometric motion of the optical star only. Their orbital periods are on the order of a year, allowing a look into another range of periods of the compact-object binaries (see, for example, \citealt[][]{saracino22}; but also \citealt{el-badry22b}).

\subsection{Caveats}

There are several caveats to our analysis, which we briefly address below.

Foremost, despite our cautious approach, the validity of the orbits is still in doubt. The sample of astrometric binaries detected by \gaia probably includes some false discoveries, as any other database would. However, as was shown by \citet{halbwachs22}, the spurious orbital solutions detected by \gaia are often characterized by high mass functions. As a result, samples of massive non-luminous companions found in the \citetalias{NSS} catalogue should be treated with some caution. Furthermore, the very nature of the \gaia astrometric 1-D measurements \citep{gaia16, pourbaix22}, the relatively small number of observations, and the fact that DR3 does not include the individual measurements imply that unambiguous detection of extremely rare systems based on DR3 data alone is challenging.
Therefore, the orbits in our sample should be validated by RV follow-up observations, for example. The amplitude of the expected modulation should be on the order of $30$ km/s, and therefore a few low-resolution observations, close to the quadrature phases, when the RVs get their extreme values, should suffice.

\red{
Second, our analysis relies on \textit{Gaia}'s reported masses, along with their uncertainty estimates \citepalias[][]{NSS}. 
Erroneous estimates of the primary masses can significantly bias the companion mass distribution. This was recently demonstrated for the NS candidate \textit{Gaia} DR3 5136025521527939072 reported by \citetalias[][]{NSS}. The primary mass of this system was probably over estimated and as a result so was that of its companion \citep[see][]{el-badry22c}.
Therefore, it will be important to use  external estimates for the stellar parameters, via spectroscopy from the LAMOST \citep{cui12} or GALAH \citep{buder21} surveys, for example.
}

Third, we emphasize that our tentative Gaussian mixture classification, separating between WD and NS candidates, is only of statistical nature. It does not account for uncertainties in the data, the prior knowledge of the physical properties, nor any additional data apart from their mass and eccentricity. To draw more specific conclusions, one might also wish to consider, for example, the spectral energy distribution, chemical composition, and Galactic trajectory of these binaries.

Looking into the future, when the next \gaia release arrives, the number of observations gets larger, and the whole astrometric data is released. Furthermore, the time span of the observations gets longer, and the sample of binaries grows substantially. We will then be able to estimate the validity of the orbits and the observational threshold for astrometric detection, deriving the statistical features of the compact-object binaries, particularly the frequency of the compact-object binaries as a function of their orbital period.

\section*{Acknowledgements}
We thank the referee, Zephyr Penoyre, for the thoughtful comments and suggestions that helped us improved the
original manuscript.
We thank Na'ama Hallakoun, Shany Danieli, Boaz Katz and Soetkin Janssens for their insightful suggestions and valuable comments. The research of SS is supported by a Benoziyo prize postdoctoral fellowship. This research was supported by Grant No.~2016069 of the United States-Israel Binational Science Foundation (BSF) and by Grant No. I-1498-303.7/2019 of the German-Israeli Foundation for Scientific Research and Development (GIF) to TM and HWR.

This work has made use of data from the European Space Agency (ESA) mission \gaia (http://www.cosmos.esa.int/gaia), processed by the Gaia Data Processing and Analysis Consortium (DPAC, http://www.cosmos.esa.int/web/gaia/dpac/consortium). Funding for the DPAC has been provided by national institutions, in particular the institutions participating in the \gaia Multilateral Agreement.

This work made use of \href{https://artpop.readthedocs.io/en/latest/index.html}{\texttt{ArtPop}}, a Python package for synthesizing stellar populations and simulating realistic images of stellar systems \citep{greco21}; the MIST isochrone grids \citep[][]{paxton11, paxton13, paxton15,choi16, dotter16}; The \href{https://github.com/SihaoCheng/WD_models}{WD models} package for WD photometry to physical parameters; \texttt{catsHTM}, a tool for fast accessing and cross-matching large astronomical catalogs \citep{soumagnac18}; \texttt{Astropy}, a community-developed core Python package for Astronomy \citep{Astropy_2013, Astropy_2018}; \texttt{matplotlib} \citep{Hunter_2007}; \texttt{numpy} \citep{Numpy_2006, Numpy_2011}; \texttt{scipy} \citep{2020SciPy-NMeth}; and \texttt{Scikit-learn} \citep{scikit-learn}.

\section*{Data Availability}

All data underlying this research are publicly available.



\bibliographystyle{mnras}
\bibliography{main}

\begin{thebibliography}{}
\makeatletter
\relax
\def\mn@urlcharsother{\let\do\@makeother \do\$\do\&\do\#\do\^\do\_\do\%\do\~}
\def\mn@doi{\begingroup\mn@urlcharsother \@ifnextchar [ {\mn@doi@}
  {\mn@doi@[]}}
\def\mn@doi@[#1]#2{\def\@tempa{#1}\ifx\@tempa\@empty \href
  {http://dx.doi.org/#2} {doi:#2}\else \href {http://dx.doi.org/#2} {#1}\fi
  \endgroup}
\def\mn@eprint#1#2{\mn@eprint@#1:#2::\@nil}
\def\mn@eprint@arXiv#1{\href {http://arxiv.org/abs/#1} {{\tt arXiv:#1}}}
\def\mn@eprint@dblp#1{\href {http://dblp.uni-trier.de/rec/bibtex/#1.xml}
  {dblp:#1}}
\def\mn@eprint@#1:#2:#3:#4\@nil{\def\@tempa {#1}\def\@tempb {#2}\def\@tempc
  {#3}\ifx \@tempc \@empty \let \@tempc \@tempb \let \@tempb \@tempa \fi \ifx
  \@tempb \@empty \def\@tempb {arXiv}\fi \@ifundefined
  {mn@eprint@\@tempb}{\@tempb:\@tempc}{\expandafter \expandafter \csname
  mn@eprint@\@tempb\endcsname \expandafter{\@tempc}}}

\bibitem[\protect\citeauthoryear{{Andrae} et~al.,}{{Andrae}
  et~al.}{2022}]{andrae22}
{Andrae} R.,  et~al., 2022, arXiv e-prints, \href
  {https://ui.adsabs.harvard.edu/abs/2022arXiv220606138A} {p. arXiv:2206.06138}

\bibitem[\protect\citeauthoryear{{Andrew}, {Penoyre}, {Belokurov}, {Evans}  \&
  {Oh}}{{Andrew} et~al.}{2022}]{andrew22}
{Andrew} S.,  {Penoyre} Z.,  {Belokurov} V.,  {Evans} N.~W.,   {Oh} S.,  2022,
  \mn@doi [\mnras] {10.1093/mnras/stac2532}, \href
  {https://ui.adsabs.harvard.edu/abs/2022MNRAS.tmp.2342A} {}

\bibitem[\protect\citeauthoryear{{Andrews} \& {Kalogera}}{{Andrews} \&
  {Kalogera}}{2022}]{andrews22a}
{Andrews} J.~J.,  {Kalogera} V.,  2022, \mn@doi [\apj]
  {10.3847/1538-4357/ac66d6}, \href
  {https://ui.adsabs.harvard.edu/abs/2022ApJ...930..159A} {930, 159}

\bibitem[\protect\citeauthoryear{{Andrews}, {Breivik}  \&
  {Chatterjee}}{{Andrews} et~al.}{2019}]{andrews19}
{Andrews} J.~J.,  {Breivik} K.,   {Chatterjee} S.,  2019, \mn@doi [\apj]
  {10.3847/1538-4357/ab441f}, \href
  {https://ui.adsabs.harvard.edu/abs/2019ApJ...886...68A} {886, 68}

\bibitem[\protect\citeauthoryear{{Andrews}, {Taggart}  \& {Foley}}{{Andrews}
  et~al.}{2022}]{andrews22b}
{Andrews} J.~J.,  {Taggart} K.,   {Foley} R.,  2022, arXiv e-prints, \href
  {https://ui.adsabs.harvard.edu/abs/2022arXiv220700680A} {p. arXiv:2207.00680}

\bibitem[\protect\citeauthoryear{{Astropy Collaboration} et~al.,}{{Astropy
  Collaboration} et~al.}{2013}]{Astropy_2013}
{Astropy Collaboration} et~al., 2013, \mn@doi [\aap]
  {10.1051/0004-6361/201322068}, \href
  {http://adsabs.harvard.edu/abs/2013A%26A...558A..33A} {558, A33}

\bibitem[\protect\citeauthoryear{{Astropy Collaboration} et~al.,}{{Astropy
  Collaboration} et~al.}{2018}]{Astropy_2018}
{Astropy Collaboration} et~al., 2018, \mn@doi [\aj] {10.3847/1538-3881/aabc4f},
  \href {http://adsabs.harvard.edu/abs/2018AJ....156..123A} {156, 123}

\bibitem[\protect\citeauthoryear{Atri et~al.,}{Atri et~al.}{2019}]{atri19}
Atri P.,  et~al., 2019, \mn@doi [Monthly Notices of the Royal Astronomical
  Society] {10.1093/mnras/stz2335}, 489, 3116

\bibitem[\protect\citeauthoryear{{Babusiaux} et~al.,}{{Babusiaux}
  et~al.}{2022}]{babusiaux22}
{Babusiaux} C.,  et~al., 2022, arXiv e-prints, \href
  {https://ui.adsabs.harvard.edu/abs/2022arXiv220605989B} {p. arXiv:2206.05989}

\bibitem[\protect\citeauthoryear{{Bashi}, {Shahaf}, {Mazeh}, {Faigler}, {Dong},
  {El-Badry}, {Rix}  \& {Jorissen}}{{Bashi} et~al.}{2022}]{bashi22}
{Bashi} D.,  {Shahaf} S.,  {Mazeh} T.,  {Faigler} S.,  {Dong} S.,  {El-Badry}
  K.,  {Rix} H.-W.,   {Jorissen} A.,  2022, arXiv e-prints, \href
  {https://ui.adsabs.harvard.edu/abs/2022arXiv220708832B} {p. arXiv:2207.08832}

\bibitem[\protect\citeauthoryear{{Beniamini} \& {Piran}}{{Beniamini} \&
  {Piran}}{2016}]{beniamini16}
{Beniamini} P.,  {Piran} T.,  2016, \mn@doi [\mnras] {10.1093/mnras/stv2903},
  \href {https://ui.adsabs.harvard.edu/abs/2016MNRAS.456.4089B} {456, 4089}

\bibitem[\protect\citeauthoryear{Benjamini \& Hochberg}{Benjamini \&
  Hochberg}{1995}]{benjamini95}
Benjamini Y.,  Hochberg Y.,  1995, Journal of the Royal statistical society:
  series B (Methodological), 57, 289

\bibitem[\protect\citeauthoryear{{Bergeron}, {Dufour}, {Fontaine}, {Coutu},
  {Blouin}, {Genest-Beaulieu}, {B{\'e}dard}  \& {Rolland}}{{Bergeron}
  et~al.}{2019}]{bergeron19}
{Bergeron} P.,  {Dufour} P.,  {Fontaine} G.,  {Coutu} S.,  {Blouin} S.,
  {Genest-Beaulieu} C.,  {B{\'e}dard} A.,   {Rolland} B.,  2019, \mn@doi [\apj]
  {10.3847/1538-4357/ab153a}, \href
  {https://ui.adsabs.harvard.edu/abs/2019ApJ...876...67B} {876, 67}

\bibitem[\protect\citeauthoryear{{Brandt} \& {Podsiadlowski}}{{Brandt} \&
  {Podsiadlowski}}{1995}]{brandt95}
{Brandt} N.,  {Podsiadlowski} P.,  1995, \mn@doi [\mnras]
  {10.1093/mnras/274.2.461}, \href
  {https://ui.adsabs.harvard.edu/abs/1995MNRAS.274..461B} {274, 461}

\bibitem[\protect\citeauthoryear{{Breivik}, {Chatterjee}  \&
  {Larson}}{{Breivik} et~al.}{2017}]{breivik17}
{Breivik} K.,  {Chatterjee} S.,   {Larson} S.~L.,  2017, \mn@doi [\apjl]
  {10.3847/2041-8213/aa97d5}, \href
  {http://adsabs.harvard.edu/abs/2017ApJ...850L..13B} {850, L13}

\bibitem[\protect\citeauthoryear{{Buder} et~al.,}{{Buder}
  et~al.}{2021}]{buder21}
{Buder} S.,  et~al., 2021, \mn@doi [\mnras] {10.1093/mnras/stab1242}, \href
  {https://ui.adsabs.harvard.edu/abs/2021MNRAS.506..150B} {506, 150}

\bibitem[\protect\citeauthoryear{{Caiazzo} et~al.,}{{Caiazzo}
  et~al.}{2021}]{caiazzo21}
{Caiazzo} I.,  et~al., 2021, \mn@doi [\nat] {10.1038/s41586-021-03615-y}, \href
  {https://ui.adsabs.harvard.edu/abs/2021Natur.595...39C} {595, 39}

\bibitem[\protect\citeauthoryear{Callister, Farr  \& Renzo}{Callister
  et~al.}{2021}]{callister21}
Callister T.~A.,  Farr W.~M.,   Renzo M.,  2021, \mn@doi [The Astrophysical
  Journal] {10.3847/1538-4357/ac1347}, 920, 157

\bibitem[\protect\citeauthoryear{{Cerda-Duran} \& {Elias-Rosa}}{{Cerda-Duran}
  \& {Elias-Rosa}}{2018}]{cerda18}
{Cerda-Duran} P.,  {Elias-Rosa} N.,  2018, in {Rezzolla} L.,  {Pizzochero} P.,
  {Jones} D.~I.,  {Rea} N.,   {Vida{\~n}a} I.,  eds,  Astrophysics and Space
  Science Library Vol. 457, Astrophysics and Space Science Library. p.~1
  (\mn@eprint {arXiv} {1806.07267}), \mn@doi{10.1007/978-3-319-97616-7_1}

\bibitem[\protect\citeauthoryear{{Chawla}, {Chatterjee}, {Breivik}, {Moorthy},
  {Andrews}  \& {Sanderson}}{{Chawla} et~al.}{2022}]{chawla22}
{Chawla} C.,  {Chatterjee} S.,  {Breivik} K.,  {Moorthy} C.~K.,  {Andrews}
  J.~J.,   {Sanderson} R.~E.,  2022, \mn@doi [\apj] {10.3847/1538-4357/ac60a5},
  \href {https://ui.adsabs.harvard.edu/abs/2022ApJ...931..107C} {931, 107}

\bibitem[\protect\citeauthoryear{{Choi}, {Dotter}, {Conroy}, {Cantiello},
  {Paxton}  \& {Johnson}}{{Choi} et~al.}{2016}]{choi16}
{Choi} J.,  {Dotter} A.,  {Conroy} C.,  {Cantiello} M.,  {Paxton} B.,
  {Johnson} B.~D.,  2016, \mn@doi [\apj] {10.3847/0004-637X/823/2/102}, \href
  {https://ui.adsabs.harvard.edu/abs/2016ApJ...823..102C} {823, 102}

\bibitem[\protect\citeauthoryear{{Creevey} et~al.,}{{Creevey}
  et~al.}{2022}]{creevey22}
{Creevey} O.~L.,  et~al., 2022, arXiv e-prints, \href
  {https://ui.adsabs.harvard.edu/abs/2022arXiv220605864C} {p. arXiv:2206.05864}

\bibitem[\protect\citeauthoryear{Cui et~al.,}{Cui et~al.}{2012}]{cui12}
Cui X.-Q.,  et~al., 2012, RAA, 12, 1197

\bibitem[\protect\citeauthoryear{Davison \& Hinkley}{Davison \&
  Hinkley}{1997}]{davison1997}
Davison A.~C.,  Hinkley D.~V.,  1997, Bootstrap Methods and their Application.
Cambridge Series in Statistical and Probabilistic Mathematics, Cambridge
  University Press, \mn@doi{10.1017/CBO9780511802843}

\bibitem[\protect\citeauthoryear{{Dewi}, {Podsiadlowski}  \& {Pols}}{{Dewi}
  et~al.}{2005}]{dewi05}
{Dewi} J.~D.~M.,  {Podsiadlowski} P.,   {Pols} O.~R.,  2005, \mn@doi [\mnras]
  {10.1111/j.1745-3933.2005.00085.x}, \href
  {https://ui.adsabs.harvard.edu/abs/2005MNRAS.363L..71D} {363, L71}

\bibitem[\protect\citeauthoryear{{Dotter}}{{Dotter}}{2016}]{dotter16}
{Dotter} A.,  2016, \mn@doi [\apjs] {10.3847/0067-0049/222/1/8}, \href
  {https://ui.adsabs.harvard.edu/abs/2016ApJS..222....8D} {222, 8}

\bibitem[\protect\citeauthoryear{{El-Badry} \& {Burdge}}{{El-Badry} \&
  {Burdge}}{2022}]{el-badry22b}
{El-Badry} K.,  {Burdge} K.~B.,  2022, \mn@doi [\mnras]
  {10.1093/mnrasl/slab135}, \href
  {https://ui.adsabs.harvard.edu/abs/2022MNRAS.511L..24E} {511, 24}

\bibitem[\protect\citeauthoryear{{El-Badry} \& {Rix}}{{El-Badry} \&
  {Rix}}{2022}]{elbadry22}
{El-Badry} K.,  {Rix} H.-W.,  2022, \mn@doi [\mnras] {10.1093/mnras/stac1797},
  \href {https://ui.adsabs.harvard.edu/abs/2022MNRAS.tmp.1730E} {}

\bibitem[\protect\citeauthoryear{{El-Badry} et~al.,}{{El-Badry}
  et~al.}{2022}]{el-badry22c}
{El-Badry} K.,  et~al., 2022, arXiv e-prints, \href
  {https://ui.adsabs.harvard.edu/abs/2022arXiv220906833E} {p. arXiv:2209.06833}

\bibitem[\protect\citeauthoryear{{Eyer} et~al.,}{{Eyer} et~al.}{2019}]{eyer19}
{Eyer} L.,  et~al., 2019, \mn@doi [\aap] {10.1051/0004-6361/201833304}, \href
  {https://ui.adsabs.harvard.edu/abs/2019A&A...623A.110G} {623, A110}

\bibitem[\protect\citeauthoryear{{Fantin} et~al.,}{{Fantin}
  et~al.}{2021}]{fantin21}
{Fantin} N.~J.,  et~al., 2021, \mn@doi [\apj] {10.3847/1538-4357/abf2b2}, \href
  {https://ui.adsabs.harvard.edu/abs/2021ApJ...913...30F} {913, 30}

\bibitem[\protect\citeauthoryear{{Fleury}, {Caiazzo}  \& {Heyl}}{{Fleury}
  et~al.}{2022a}]{fleury22a}
{Fleury} L.,  {Caiazzo} I.,   {Heyl} J.,  2022a, arXiv e-prints, \href
  {https://ui.adsabs.harvard.edu/abs/2022arXiv220501015F} {p. arXiv:2205.01015}

\bibitem[\protect\citeauthoryear{{Fleury}, {Caiazzo}  \& {Heyl}}{{Fleury}
  et~al.}{2022b}]{fleury22b}
{Fleury} L.,  {Caiazzo} I.,   {Heyl} J.,  2022b, \mn@doi [\mnras]
  {10.1093/mnras/stac458}, \href
  {https://ui.adsabs.harvard.edu/abs/2022MNRAS.511.5984F} {511, 5984}

\bibitem[\protect\citeauthoryear{{Fryer}, {Belczynski}, {Wiktorowicz},
  {Dominik}, {Kalogera}  \& {Holz}}{{Fryer} et~al.}{2012}]{fryer12}
{Fryer} C.~L.,  {Belczynski} K.,  {Wiktorowicz} G.,  {Dominik} M.,  {Kalogera}
  V.,   {Holz} D.~E.,  2012, \mn@doi [\apj] {10.1088/0004-637X/749/1/91}, \href
  {https://ui.adsabs.harvard.edu/abs/2012ApJ...749...91F} {749, 91}

\bibitem[\protect\citeauthoryear{{Gaia Collaboration} et~al.,}{{Gaia
  Collaboration} et~al.}{2016}]{gaia16}
{Gaia Collaboration} et~al., 2016, \mn@doi [\aap]
  {10.1051/0004-6361/201629272}, \href
  {https://ui.adsabs.harvard.edu/abs/2016A&A...595A...1G} {595, A1}

\bibitem[\protect\citeauthoryear{{Gaia Collaboration} et~al.,}{{Gaia
  Collaboration} et~al.}{2022}]{NSS}
{Gaia Collaboration} et~al., 2022, arXiv e-prints, \href
  {https://ui.adsabs.harvard.edu/abs/2022arXiv220605595G} {p. arXiv:2206.05595}

\bibitem[\protect\citeauthoryear{{Gentile Fusillo} et~al.,}{{Gentile Fusillo}
  et~al.}{2021}]{GentileFusillo21}
{Gentile Fusillo} N.~P.,  et~al., 2021, \mn@doi [\mnras]
  {10.1093/mnras/stab2672}, \href
  {https://ui.adsabs.harvard.edu/abs/2021MNRAS.508.3877G} {508, 3877}

\bibitem[\protect\citeauthoryear{{Greco} \& {Danieli}}{{Greco} \&
  {Danieli}}{2021}]{greco21}
{Greco} J.~P.,  {Danieli} S.,  2021, arXiv e-prints, \href
  {https://ui.adsabs.harvard.edu/abs/2021arXiv210913943G} {p. arXiv:2109.13943}

\bibitem[\protect\citeauthoryear{{Halbwachs} et~al.,}{{Halbwachs}
  et~al.}{2022}]{halbwachs22}
{Halbwachs} J.-L.,  et~al., 2022, arXiv e-prints, \href
  {https://ui.adsabs.harvard.edu/abs/2022arXiv220605726H} {p. arXiv:2206.05726}

\bibitem[\protect\citeauthoryear{Hansen \& Phinney}{Hansen \&
  Phinney}{1997}]{hansen97}
Hansen B. M.~S.,  Phinney E.~S.,  1997, \mn@doi [Monthly Notices of the Royal
  Astronomical Society] {10.1093/mnras/291.3.569}, 291, 569

\bibitem[\protect\citeauthoryear{{Heacox}}{{Heacox}}{1995}]{heacox95}
{Heacox} W.~D.,  1995, \mn@doi [\aj] {10.1086/117480}, \href
  {https://ui.adsabs.harvard.edu/abs/1995AJ....109.2670H} {109, 2670}

\bibitem[\protect\citeauthoryear{{Heger}, {Fryer}, {Woosley}, {Langer}  \&
  {Hartmann}}{{Heger} et~al.}{2003}]{heger03}
{Heger} A.,  {Fryer} C.~L.,  {Woosley} S.~E.,  {Langer} N.,   {Hartmann} D.~H.,
   2003, \mn@doi [\apj] {10.1086/375341}, \href
  {https://ui.adsabs.harvard.edu/abs/2003ApJ...591..288H} {591, 288}

\bibitem[\protect\citeauthoryear{{Heintz}, {Hermes}, {El-Badry}, {Walsh}, {van
  Saders}, {Fields}  \& {Koester}}{{Heintz} et~al.}{2022}]{heintz22}
{Heintz} T.~M.,  {Hermes} J.~J.,  {El-Badry} K.,  {Walsh} C.,  {van Saders}
  J.~L.,  {Fields} C.~E.,   {Koester} D.,  2022, \mn@doi [\apj]
  {10.3847/1538-4357/ac78d9}, \href
  {https://ui.adsabs.harvard.edu/abs/2022ApJ...934..148H} {934, 148}

\bibitem[\protect\citeauthoryear{{Hills}}{{Hills}}{1983}]{hills83}
{Hills} J.~G.,  1983, \mn@doi [\apj] {10.1086/160871}, \href
  {https://ui.adsabs.harvard.edu/abs/1983ApJ...267..322H} {267, 322}

\bibitem[\protect\citeauthoryear{{Hollands}, {Tremblay}, {G{\"a}nsicke},
  {Gentile-Fusillo}  \& {Toonen}}{{Hollands} et~al.}{2018}]{hollands18}
{Hollands} M.~A.,  {Tremblay} P.~E.,  {G{\"a}nsicke} B.~T.,  {Gentile-Fusillo}
  N.~P.,   {Toonen} S.,  2018, \mn@doi [\mnras] {10.1093/mnras/sty2057}, \href
  {https://ui.adsabs.harvard.edu/abs/2018MNRAS.480.3942H} {480, 3942}

\bibitem[\protect\citeauthoryear{Hunter}{Hunter}{2007}]{Hunter_2007}
Hunter J.~D.,  2007, \mn@doi [Computing In Science \& Engineering]
  {10.1109/MCSE.2007.55}, 9, 90

\bibitem[\protect\citeauthoryear{Igoshev \& Perets}{Igoshev \&
  Perets}{2019}]{igoshev19}
Igoshev A.~P.,  Perets H.~B.,  2019, \mn@doi [Monthly Notices of the Royal
  Astronomical Society] {10.1093/mnras/stz1024}, 486, 4098

\bibitem[\protect\citeauthoryear{{Izzard}, {Dermine}  \& {Church}}{{Izzard}
  et~al.}{2010}]{izzard10}
{Izzard} R.~G.,  {Dermine} T.,   {Church} R.~P.,  2010, \mn@doi [\aap]
  {10.1051/0004-6361/201015254}, \href
  {https://ui.adsabs.harvard.edu/abs/2010A&A...523A..10I} {523, A10}

\bibitem[\protect\citeauthoryear{{Janssens} et~al.,}{{Janssens}
  et~al.}{2022}]{janssens22}
{Janssens} S.,  et~al., 2022, \mn@doi [\aap] {10.1051/0004-6361/202141866},
  \href {https://ui.adsabs.harvard.edu/abs/2022A&A...658A.129J} {658, A129}

\bibitem[\protect\citeauthoryear{{Jorissen} \& {Frankowski}}{{Jorissen} \&
  {Frankowski}}{2008}]{jorissen08}
{Jorissen} A.,  {Frankowski} A.,  2008, in {Pellegrini} P.,  {Daflon} S.,
  {Alcaniz} J.~S.,   {Telles} E.,  eds,  American Institute of Physics
  Conference Series Vol. 1057, Graduate School in Astronomy: XII Special
  Courses at the National Observatory of Rio de Janeiro. pp 1--55 (\mn@eprint
  {arXiv} {0804.3720}), \mn@doi{10.1063/1.2999998}

\bibitem[\protect\citeauthoryear{{Jorissen}, {Boffin}, {Karinkuzhi}, {Van Eck},
  {Escorza}, {Shetye}  \& {Van Winckel}}{{Jorissen} et~al.}{2019}]{jorissen19}
{Jorissen} A.,  {Boffin} H.~M.~J.,  {Karinkuzhi} D.,  {Van Eck} S.,  {Escorza}
  A.,  {Shetye} S.,   {Van Winckel} H.,  2019, \mn@doi [\aap]
  {10.1051/0004-6361/201834630}, \href
  {https://ui.adsabs.harvard.edu/abs/2019A&A...626A.127J} {626, A127}

\bibitem[\protect\citeauthoryear{{Kalogera}}{{Kalogera}}{1996}]{kalogera96}
{Kalogera} V.,  1996, \mn@doi [\apj] {10.1086/177974}, \href
  {https://ui.adsabs.harvard.edu/abs/1996ApJ...471..352K} {471, 352}

\bibitem[\protect\citeauthoryear{{Kilic}, {Bergeron}, {Blouin}  \&
  {B{\'e}dard}}{{Kilic} et~al.}{2021}]{kilic21}
{Kilic} M.,  {Bergeron} P.,  {Blouin} S.,   {B{\'e}dard} A.,  2021, \mn@doi
  [\mnras] {10.1093/mnras/stab767}, \href
  {https://ui.adsabs.harvard.edu/abs/2021MNRAS.503.5397K} {503, 5397}

\bibitem[\protect\citeauthoryear{{Knigge}, {Baraffe}  \& {Patterson}}{{Knigge}
  et~al.}{2011}]{CV11}
{Knigge} C.,  {Baraffe} I.,   {Patterson} J.,  2011, \mn@doi [\apjs]
  {10.1088/0067-0049/194/2/28}, \href
  {https://ui.adsabs.harvard.edu/abs/2011ApJS..194...28K} {194, 28}

\bibitem[\protect\citeauthoryear{{Kreidberg}, {Bailyn}, {Farr}  \&
  {Kalogera}}{{Kreidberg} et~al.}{2012}]{MassGap12}
{Kreidberg} L.,  {Bailyn} C.~D.,  {Farr} W.~M.,   {Kalogera} V.,  2012, \mn@doi
  [\apj] {10.1088/0004-637X/757/1/36}, \href
  {https://ui.adsabs.harvard.edu/abs/2012ApJ...757...36K} {757, 36}

\bibitem[\protect\citeauthoryear{{Lam} et~al.,}{{Lam}
  et~al.}{2022}]{MassGap22a}
{Lam} C.~Y.,  et~al., 2022, \mn@doi [\apjl] {10.3847/2041-8213/ac7442}, \href
  {https://ui.adsabs.harvard.edu/abs/2022ApJ...933L..23L} {933, L23}

\bibitem[\protect\citeauthoryear{{Lattimer} \& {Steiner}}{{Lattimer} \&
  {Steiner}}{2014}]{lattimer14}
{Lattimer} J.~M.,  {Steiner} A.~W.,  2014, \mn@doi [\apj]
  {10.1088/0004-637X/784/2/123}, \href
  {https://ui.adsabs.harvard.edu/abs/2014ApJ...784..123L} {784, 123}

\bibitem[\protect\citeauthoryear{{Manchester}}{{Manchester}}{2017}]{manchester17}
{Manchester} R.~N.,  2017, \mn@doi [Journal of Astrophysics and Astronomy]
  {10.1007/s12036-017-9469-2}, \href
  {https://ui.adsabs.harvard.edu/abs/2017JApA...38...42M} {38, 42}

\bibitem[\protect\citeauthoryear{{Martinez} et~al.,}{{Martinez}
  et~al.}{2015}]{martinez15}
{Martinez} J.~G.,  et~al., 2015, \mn@doi [\apj] {10.1088/0004-637X/812/2/143},
  \href {https://ui.adsabs.harvard.edu/abs/2015ApJ...812..143M} {812, 143}

\bibitem[\protect\citeauthoryear{{Mashian} \& {Loeb}}{{Mashian} \&
  {Loeb}}{2017}]{mashian17}
{Mashian} N.,  {Loeb} A.,  2017, \mn@doi [\mnras] {10.1093/mnras/stx1410},
  \href {https://ui.adsabs.harvard.edu/abs/2017MNRAS.470.2611M} {470, 2611}

\bibitem[\protect\citeauthoryear{{Mazeh} et~al.,}{{Mazeh}
  et~al.}{2022}]{Mazeh22}
{Mazeh} T.,  et~al., 2022, arXiv e-prints, \href
  {https://ui.adsabs.harvard.edu/abs/2022arXiv220611270M} {p. arXiv:2206.11270}

\bibitem[\protect\citeauthoryear{{Miller}, {Caiazzo}, {Heyl}, {Richer}  \&
  {Tremblay}}{{Miller} et~al.}{2022}]{miller22}
{Miller} D.~R.,  {Caiazzo} I.,  {Heyl} J.,  {Richer} H.~B.,   {Tremblay} P.-E.,
   2022, \mn@doi [\apjl] {10.3847/2041-8213/ac50a5}, \href
  {https://ui.adsabs.harvard.edu/abs/2022ApJ...926L..24M} {926, L24}

\bibitem[\protect\citeauthoryear{{Morrell} \& {Naylor}}{{Morrell} \&
  {Naylor}}{2019}]{morrell19}
{Morrell} S.,  {Naylor} T.,  2019, \mn@doi [\mnras] {10.1093/mnras/stz2242},
  \href {https://ui.adsabs.harvard.edu/abs/2019MNRAS.489.2615M} {489, 2615}

\bibitem[\protect\citeauthoryear{Oliphant}{Oliphant}{2006}]{Numpy_2006}
Oliphant T.,  2006, {NumPy}: A guide to {NumPy}, USA: Trelgol Publishing, \url
  {http://www.numpy.org/}

\bibitem[\protect\citeauthoryear{{{\"O}zel} \& {Freire}}{{{\"O}zel} \&
  {Freire}}{2016}]{ozel16}
{{\"O}zel} F.,  {Freire} P.,  2016, \mn@doi [\araa]
  {10.1146/annurev-astro-081915-023322}, \href
  {https://ui.adsabs.harvard.edu/abs/2016ARA&A..54..401O} {54, 401}

\bibitem[\protect\citeauthoryear{{Paul}}{{Paul}}{2017}]{paul17}
{Paul} B.,  2017, \mn@doi [Journal of Astrophysics and Astronomy]
  {10.1007/s12036-017-9475-4}, \href
  {https://ui.adsabs.harvard.edu/abs/2017JApA...38...39P} {38, 39}

\bibitem[\protect\citeauthoryear{{Paxton}, {Bildsten}, {Dotter}, {Herwig},
  {Lesaffre}  \& {Timmes}}{{Paxton} et~al.}{2011}]{paxton11}
{Paxton} B.,  {Bildsten} L.,  {Dotter} A.,  {Herwig} F.,  {Lesaffre} P.,
  {Timmes} F.,  2011, \mn@doi [\apjs] {10.1088/0067-0049/192/1/3}, \href
  {https://ui.adsabs.harvard.edu/abs/2011ApJS..192....3P} {192, 3}

\bibitem[\protect\citeauthoryear{{Paxton} et~al.,}{{Paxton}
  et~al.}{2013}]{paxton13}
{Paxton} B.,  et~al., 2013, \mn@doi [\apjs] {10.1088/0067-0049/208/1/4}, \href
  {https://ui.adsabs.harvard.edu/abs/2013ApJS..208....4P} {208, 4}

\bibitem[\protect\citeauthoryear{{Paxton} et~al.,}{{Paxton}
  et~al.}{2015}]{paxton15}
{Paxton} B.,  et~al., 2015, \mn@doi [\apjs] {10.1088/0067-0049/220/1/15}, \href
  {https://ui.adsabs.harvard.edu/abs/2015ApJS..220...15P} {220, 15}

\bibitem[\protect\citeauthoryear{{Pecaut} \& {Mamajek}}{{Pecaut} \&
  {Mamajek}}{2013}]{mamajek13}
{Pecaut} M.~J.,  {Mamajek} E.~E.,  2013, \mn@doi [\apjs]
  {10.1088/0067-0049/208/1/9}, \href
  {https://ui.adsabs.harvard.edu/abs/2013ApJS..208....9P} {208, 9}

\bibitem[\protect\citeauthoryear{Pedregosa et~al.,}{Pedregosa
  et~al.}{2011}]{scikit-learn}
Pedregosa F.,  et~al., 2011, Journal of Machine Learning Research, 12, 2825

\bibitem[\protect\citeauthoryear{{Penoyre}, {Belokurov}  \& {Evans}}{{Penoyre}
  et~al.}{2022}]{penoyre22}
{Penoyre} Z.,  {Belokurov} V.,   {Evans} N.~W.,  2022, \mn@doi [\mnras]
  {10.1093/mnras/stac959}, \href
  {https://ui.adsabs.harvard.edu/abs/2022MNRAS.513.2437P} {513, 2437}

\bibitem[\protect\citeauthoryear{{Pfahl}, {Rappaport}, {Podsiadlowski}  \&
  {Spruit}}{{Pfahl} et~al.}{2002}]{pfahl02}
{Pfahl} E.,  {Rappaport} S.,  {Podsiadlowski} P.,   {Spruit} H.,  2002, \mn@doi
  [\apj] {10.1086/340794}, \href
  {https://ui.adsabs.harvard.edu/abs/2002ApJ...574..364P} {574, 364}

\bibitem[\protect\citeauthoryear{{Pourbaix} et~al.,}{{Pourbaix}
  et~al.}{2022}]{pourbaix22}
{Pourbaix} D.,  et~al., 2022, {Gaia DR3 documentation Chapter 7: Non-single
  stars}, Gaia DR3 documentation, European Space Agency; Gaia Data Processing
  and Analysis Consortium.

\bibitem[\protect\citeauthoryear{{Robinson}}{{Robinson}}{1976}]{CV76}
{Robinson} E.~L.,  1976, \mn@doi [\araa] {10.1146/annurev.aa.14.090176.001003},
  \href {https://ui.adsabs.harvard.edu/abs/1976ARA&A..14..119R} {14, 119}

\bibitem[\protect\citeauthoryear{Rousseeuw}{Rousseeuw}{1987}]{rousseeuw87}
Rousseeuw P.~J.,  1987, \mn@doi [Journal of Computational and Applied
  Mathematics] {https://doi.org/10.1016/0377-0427(87)90125-7}, 20, 53

\bibitem[\protect\citeauthoryear{{Saracino} et~al.,}{{Saracino}
  et~al.}{2022}]{saracino22}
{Saracino} S.,  et~al., 2022, \mn@doi [\mnras] {10.1093/mnras/stab3159}, \href
  {https://ui.adsabs.harvard.edu/abs/2022MNRAS.511.2914S} {511, 2914}

\bibitem[\protect\citeauthoryear{{Shahaf}, {Mazeh}  \& {Faigler}}{{Shahaf}
  et~al.}{2017}]{shahaf17}
{Shahaf} S.,  {Mazeh} T.,   {Faigler} S.,  2017, \mn@doi [\mnras]
  {10.1093/mnras/stx2257}, \href
  {https://ui.adsabs.harvard.edu/abs/2017MNRAS.472.4497S} {472, 4497}

\bibitem[\protect\citeauthoryear{{Shahaf}, {Mazeh}, {Faigler}  \&
  {Holl}}{{Shahaf} et~al.}{2019}]{shahaf19}
{Shahaf} S.,  {Mazeh} T.,  {Faigler} S.,   {Holl} B.,  2019, \mn@doi [\mnras]
  {10.1093/mnras/stz1636}, \href
  {https://ui.adsabs.harvard.edu/abs/2019MNRAS.487.5610S} {487, 5610}

\bibitem[\protect\citeauthoryear{{Shenar} et~al.,}{{Shenar}
  et~al.}{2022a}]{shenar22}
{Shenar} T.,  et~al., 2022a, \mn@doi [Nature Astronomy]
  {10.1038/s41550-022-01730-y}, \href
  {https://ui.adsabs.harvard.edu/abs/2022NatAs.tmp..171S} {}

\bibitem[\protect\citeauthoryear{{Shenar} et~al.,}{{Shenar}
  et~al.}{2022b}]{shenar22b}
{Shenar} T.,  et~al., 2022b, arXiv e-prints, \href
  {https://ui.adsabs.harvard.edu/abs/2022arXiv220707674S} {p. arXiv:2207.07674}

\bibitem[\protect\citeauthoryear{{Smart} et~al.,}{{Smart}
  et~al.}{2021}]{smart21}
{Smart} R.~L.,  et~al., 2021, \mn@doi [\aap] {10.1051/0004-6361/202039498},
  \href {https://ui.adsabs.harvard.edu/abs/2021A&A...649A...6G} {649, A6}

\bibitem[\protect\citeauthoryear{{Soumagnac} \& {Ofek}}{{Soumagnac} \&
  {Ofek}}{2018}]{soumagnac18}
{Soumagnac} M.~T.,  {Ofek} E.~O.,  2018, \mn@doi [\pasp]
  {10.1088/1538-3873/aac410}, \href
  {https://ui.adsabs.harvard.edu/abs/2018PASP..130g5002S} {130, 075002}

\bibitem[\protect\citeauthoryear{{Tauris} et~al.,}{{Tauris}
  et~al.}{2017}]{tauris17}
{Tauris} T.~M.,  et~al., 2017, \mn@doi [\apj] {10.3847/1538-4357/aa7e89}, \href
  {https://ui.adsabs.harvard.edu/abs/2017ApJ...846..170T} {846, 170}

\bibitem[\protect\citeauthoryear{{Toonen}, {Claeys}, {Mennekens}  \&
  {Ruiter}}{{Toonen} et~al.}{2014}]{toonen14}
{Toonen} S.,  {Claeys} J.~S.~W.,  {Mennekens} N.,   {Ruiter} A.~J.,  2014,
  \mn@doi [\aap] {10.1051/0004-6361/201321576}, \href
  {https://ui.adsabs.harvard.edu/abs/2014A&A...562A..14T} {562, A14}

\bibitem[\protect\citeauthoryear{{Torres}, {Rebassa-Mansergas}, {Camisassa}  \&
  {Raddi}}{{Torres} et~al.}{2021}]{torres21}
{Torres} S.,  {Rebassa-Mansergas} A.,  {Camisassa} M.~E.,   {Raddi} R.,  2021,
  \mn@doi [\mnras] {10.1093/mnras/stab079}, \href
  {https://ui.adsabs.harvard.edu/abs/2021MNRAS.502.1753T} {502, 1753}

\bibitem[\protect\citeauthoryear{{Tremblay}, {Cummings}, {Kalirai},
  {G{\"a}nsicke}, {Gentile-Fusillo}  \& {Raddi}}{{Tremblay}
  et~al.}{2016}]{tremblay16}
{Tremblay} P.~E.,  {Cummings} J.,  {Kalirai} J.~S.,  {G{\"a}nsicke} B.~T.,
  {Gentile-Fusillo} N.,   {Raddi} R.,  2016, \mn@doi [\mnras]
  {10.1093/mnras/stw1447}, \href
  {https://ui.adsabs.harvard.edu/abs/2016MNRAS.461.2100T} {461, 2100}

\bibitem[\protect\citeauthoryear{{Van der Swaelmen}, {Boffin}, {Jorissen}  \&
  {Van Eck}}{{Van der Swaelmen} et~al.}{2017}]{swaelmen17}
{Van der Swaelmen} M.,  {Boffin} H.~M.~J.,  {Jorissen} A.,   {Van Eck} S.,
  2017, \mn@doi [\aap] {10.1051/0004-6361/201628867}, \href
  {https://ui.adsabs.harvard.edu/abs/2017A&A...597A..68V} {597, A68}

\bibitem[\protect\citeauthoryear{Virtanen et~al.,}{Virtanen
  et~al.}{2020}]{2020SciPy-NMeth}
Virtanen P.,  et~al., 2020, \mn@doi [Nature Methods]
  {10.1038/s41592-019-0686-2}, \href {https://rdcu.be/b08Wh} {17, 261}

\bibitem[\protect\citeauthoryear{{\VAN{Walt}{Van der}{van der} Walt}, {Colbert}
   \& {Varoquaux}}{{\VAN{Walt}{Van der}{van der} Walt}
  et~al.}{2011}]{Numpy_2011}
{\VAN{Walt}{Van der}{van der} Walt} S.,  {Colbert} S.~C.,   {Varoquaux} G.,
  2011, \mn@doi [Computing in Science Engineering] {10.1109/MCSE.2011.37}, 13,
  22

\bibitem[\protect\citeauthoryear{Willcox, Mandel, Thrane, Deller, Stevenson  \&
  Vigna-Gómez}{Willcox et~al.}{2021}]{willcox21}
Willcox R.,  Mandel I.,  Thrane E.,  Deller A.,  Stevenson S.,   Vigna-Gómez
  A.,  2021, \mn@doi [The Astrophysical Journal Letters]
  {10.3847/2041-8213/ac2cc8}, 920, L37

\bibitem[\protect\citeauthoryear{{Yamaguchi}, {Kawanaka}, {Bulik}  \&
  {Piran}}{{Yamaguchi} et~al.}{2018}]{yamaguchi21}
{Yamaguchi} M.~S.,  {Kawanaka} N.,  {Bulik} T.,   {Piran} T.,  2018, \mn@doi
  [\apj] {10.3847/1538-4357/aac5ec}, \href
  {https://ui.adsabs.harvard.edu/abs/2018ApJ...861...21Y} {861, 21}

\bibitem[\protect\citeauthoryear{{Ye} \& {Fishbach}}{{Ye} \&
  {Fishbach}}{2022}]{MassGap22b}
{Ye} C.,  {Fishbach} M.,  2022, arXiv e-prints, \href
  {https://ui.adsabs.harvard.edu/abs/2022arXiv220205164Y} {p. arXiv:2202.05164}

\bibitem[\protect\citeauthoryear{{Zahn}}{{Zahn}}{1977}]{zahn77}
{Zahn} J.~P.,  1977, \aap, \href
  {https://ui.adsabs.harvard.edu/abs/1977A&A....57..383Z} {57, 383}

\bibitem[\protect\citeauthoryear{{van de Kamp}}{{van de Kamp}}{1975}]{kamp75}
{van de Kamp} P.,  1975, \mn@doi [\araa] {10.1146/annurev.aa.13.090175.001455},
  \href {https://ui.adsabs.harvard.edu/abs/1975ARA&A..13..295V} {13, 295}

\bibitem[\protect\citeauthoryear{{van den Heuvel}}{{van den
  Heuvel}}{2007}]{heuvel07}
{van den Heuvel} E.~P.~J.,  2007, in {di Salvo} T.,  {Israel} G.~L.,
  {Piersant} L.,  {Burderi} L.,  {Matt} G.,  {Tornambe} A.,   {Menna} M.~T.,
  eds,  American Institute of Physics Conference Series Vol. 924, The
  Multicolored Landscape of Compact Objects and Their Explosive Origins. pp
  598--606 (\mn@eprint {arXiv} {0704.1215}), \mn@doi{10.1063/1.2774916}

\makeatother
\end{thebibliography}



\bsp	
\label{lastpage}
\end{document}